\documentclass[aps,prb,twocolumn]{revtex4-2}

\usepackage{amsmath}
\usepackage{amsfonts}
\usepackage{amssymb}
\usepackage{graphicx}
\usepackage{tabularx}
\usepackage{color}

\begin{document}
\title{Thermally-driven phase transitions in freestanding\\low-buckled silicene, germanene, and stanene}
\author{John M. Davis}
\affiliation{Department of Physics, University of Arkansas, Fayetteville, AR 72701, USA}
\author{Gustavo Orozco-Galvan}
\affiliation{Department of Physics, University of Arkansas, Fayetteville, AR 72701, USA}
\author{Salvador Barraza-Lopez}
\email{sbarraza@uark.edu}
\affiliation{Department of Physics, University of Arkansas, Fayetteville, AR 72701, USA}
\begin{abstract}
Low-buckled silicene, germanene, and stanene are group$-IV$ graphene allotropes. They form a honeycomb lattice out of two interpenetrating ($A$ and $B$) triangular sublattices that are vertically separated by a small distance $\Delta_z$. The atomic numbers $Z$ of silicon, germanium, and tin are larger to carbon's ($Z_C=6$), making them the first experimentally viable two-dimensional topological insulators. Those materials have a twice-energy-degenerate atomistic structure characterized by the buckling direction of the $B$ sublattice with respect to the $A$ sublattice [whereby the $B-$atom either protrudes {\em above} ($\Delta_z>0$) or {\em below} ($\Delta_z<0$) the $A-$atoms], and the consequences of that energy degeneracy on their elastic and electronic properties have not been reported thus far. Here, we uncover {\em ferroelastic, bistable} behavior on silicene, which turns into an {\em average} planar structure at about 600 K. Further, the creation of electron and hole puddles obfuscates the zero-temperature SOC induced band gaps at temperatures as low as 200 K, which may discard silicene as a viable two-dimensional topological insulator for room temperature applications. Germanene, on the other hand, never undergoes a low-buckled to planar 2D transformation, becoming amorphous at around 675 K instead, and preserving its SOC-induced bandgap despite of band broadening. Stanene undergoes a transition onto a crystalline 3D structure at about 300 K, preserving its SOC-induced electronic band gap up to that temperature. Unlike what is observed in silicene and germanene, stanene readily develops a higher-coordinated structure with a high degree of structural order. The structural phenomena is shown to have deep-reaching consequences for the electronic and vibrational properties of those two dimensional topological insulators.
\end{abstract}
\maketitle

\section{Introduction}

Studies of the topology of the electronic structure originated on two-dimensional Kekul\'e \cite{Haldane} and hexagonal lattices \cite{Kane}.
Spin-orbit coupling plays a key role on the topology of the electronic band structure \cite{shen2013topological}, and the earliest studies of the effect of spin-orbit coupling (SOC) on a hexagonal lattice (graphene) are due to Huertas-Hernando, Guinea, and Brataas \cite{Huertas-Hernando}, and to Kane and Mele \cite{Kane}. The latter work emphasized the change of topology of electrons in the valence and conduction bands due to SOC. The SOC strength increases  with the atomic number $Z$ as $Z^4$ \cite{RevModPhys.77.1375}, and it can also be tuned by the curvature induced the two atoms in the unit cell (u.c.) are at a relative height $\Delta_z\ne 0$ \cite{Huertas-Hernando}. (The word low-buckled was first introduced in Ref.~\cite{Ciraci} which published after Ref.~\cite{Huertas-Hernando}, and it provides a conventional name for those structures). Graphene's SOC is small because of carbon's small $Z_C=6$, and it is larger on honeycomb lattices containing heavier elements.

Proposals for graphene analogs with strong SOC began appeared early on: low-buckled silicene and germanene's were first studied in 1994 \cite{Takeda}, and low-buckled stanene in 2011 \cite{stanene1}. Nevertheless, germanene and stanene possess what we called a ``high-buckled'' two-dimensional structural ground state structure with nine-fold coordination (a hexagonal closed-packed bilayer) on their freestanding form \cite{bib:rivero}. The high-buckled phase for stanene has been suggested to be a two-dimensional superconductor \cite{superconductivity}. These two-dimensional materials continue to be explored, with copious experimental work on growth and characterization being reported~\cite{Ciraci,Lede,Carey,stanene2,Zhang,Kara1,siliceneExpt,GermaneneRecent1,GermaneneRecent,exptStanene,Dong_2021}, as well as multiple reviews on the subject \cite{revSilicene1,OUGHADDOU201546,revGermanene,revGermanene2,revStanene,revStanene2}.

There is still one intrinsic physical quality of freestanding, low-buckled silicene, germanene, and stanene eluding attention though: namely, their propensity to undergo structural phase transformations, which necessarily influence their electronic and vibrational properties. Such study is the topic of this work. In fact, the buckled structure of silicene, germanene, and stanene gives a reason to expect structural transformations: the atomistic energy remains the same upon a mirror reflection with respect to the two-dimensional plane, making these materials structurally degenerate. Structural degeneracies have been found to give rise to structural phase transitions \cite{Mehboudi2016,Mehboudi2016b,prb2018,Tyler,editorial,Thermoelectric,review,Juan}, to non-harmonic phonon modes \cite{Delaire}, and to softened elastic constants \cite{Alejandro,Joe2022}.

The manuscript is structured as follows: a description of computational methods is provided in Sec.~\ref{sec:methods}. The structural energy of low- and high-buckled silicene, germanene, and stanene is then studied as a function of the two structural parameters (the lattice constant $a_0$ and the buckling height $\Delta_z$) at zero-temperature in Sec.~\ref{sec:ii}. Such energy information then guides a finite-temperature study of those materials (Section \ref{sec:MD}) in which the structural, vibrational, and electronic properties are comparatively studied. Conclusions are presented in Sec.~\ref{sec:conclusion}.

\section{Computational Methods}\label{sec:methods}

The study is based on density-functional theory \cite{martin_2004} as implemented in the {\em SIESTA} code \cite{bib:siesta}, with in-house tuned pseudopotentials \cite{RIVERO2015372} in the generalized gradient approximation for the exchange-correlation functional due to Perdew, Burke, and Ernzerhof \cite{bib:gga}.

\subsection{Zero-temperature calculations}
Zero-temperature calculations were first carried out to understand the elastic energy landscape and vibrational properties of low-buckled structures within the harmonic approximation. Zero-temperature phonon calculations within the harmonic approximation utilized a $9\times 15$ rectangular supercell and atomic displacements of either 1.0 Bohr (silicene and germanene) or 0.9 Bohr (stanene).

\subsection{Finite-temperature calculations}
{\em Ab initio} molecular dynamics (MD) calculations within the isothermal-isobaric (NPT) ensemble were carried out at standard ambient pressure. Temperature was controlled by a Nos\'e thermostat, while a target pressure was set using the Parinello-Raman method \cite{MD}. We opted away from a supercell having a diamond shape for one with a configuration close to a square, and set up a 7$\times$12 supercell using a non-primitive rectangular cell containing four atoms and in-plane lattice vectors $\textbf{a}_1=a_{0,lb}(\sqrt{3},0,0)$ and $\textbf{a}_2=a_{0,lb}(0,1,0)$ [with $a_{0,lb}$ the lattice constant of the primitive unit cell (u.c.) at zero temperature], which correspond to the armchair and zigzag directions, respectively. The 7$\times$12 supercells we built have an almost one-to-one horizontal/vertical aspect ratio, spanning an area $7\sqrt{3}a_0\times 12a_0\simeq 12.12 a_0\times 12 a_0$, and contain 336 atoms. We initialized atomic positions in the low-buckled two-dimensional silicene, germanene, or stanene structure shown on Fig.~\ref{fig:1}(b)(i). The out-of-plane lattice vector had a length of 15 \AA{}.

MD calculations were carried out at temperatures ranging from 200 K to 900 K for silicene, 100 K to 675 K for germanene, and 100 K to 400 K for stanene. A MD time step length of 1.5 fs was used, unless indicated otherwise. The instantaneous temperature, the energy per primitive u.~c., as well as $a_{0}$ and $\Delta_z$ were tracked. $\Delta_z$ is a signed height obtained with the procedure given in Appendix \ref{sec:appendixA}. Pair-correlation functions $g(r)$ were computed as well, to hint of structural transitions onto three-dimensional or disordered phases. Those were obtained  for all frames after 20 ps, by counting the number of neighboring atoms located at a distance $r$ at a given value in between 0 \AA{} to $5a_0$, with a $5a_0\times 10^{-3}$ resolution. The count was normalized by $4\pi r^2$.

We also employed a {\em non-perturbative} method to obtain the natural vibrational frequencies of these materials at finite temperature, relying on the power spectrum of the velocity autocorrelation function within MD trajectories. The power spectrum is obtained at a discrete set of $7\times 12$ $k-$points by (i) Fourier transforming atomic velocities, (ii) carrying out a time autocorrelation, and (iii) Fourier transforming the resulting function into frequency space \cite{bib:koukaras}. The process has been employed to determine finite-temperature behavior of ferroic 2D materials, in which softening of specific vibrational modes has been observed \cite{Tyler,Thermoelectric}.

The electronic structure was calculated on one hundred primitive u.c.s. Each u.c.~contains the average atomic positions of a supercell at a fixed time during the structural time-evolution. The use of multiple unit cells permits capturing structural fluctuations as time evolves, and each bandstructure at a given time is independently plotted to make such fluctuations observable. Those structures were obtained at times equally spaced within 300 fs after thermal equilibration.

\section{Atomistic structure and energy degeneracies}\label{sec:ii}

\begin{figure}[tb]
\includegraphics[width=0.48\textwidth]{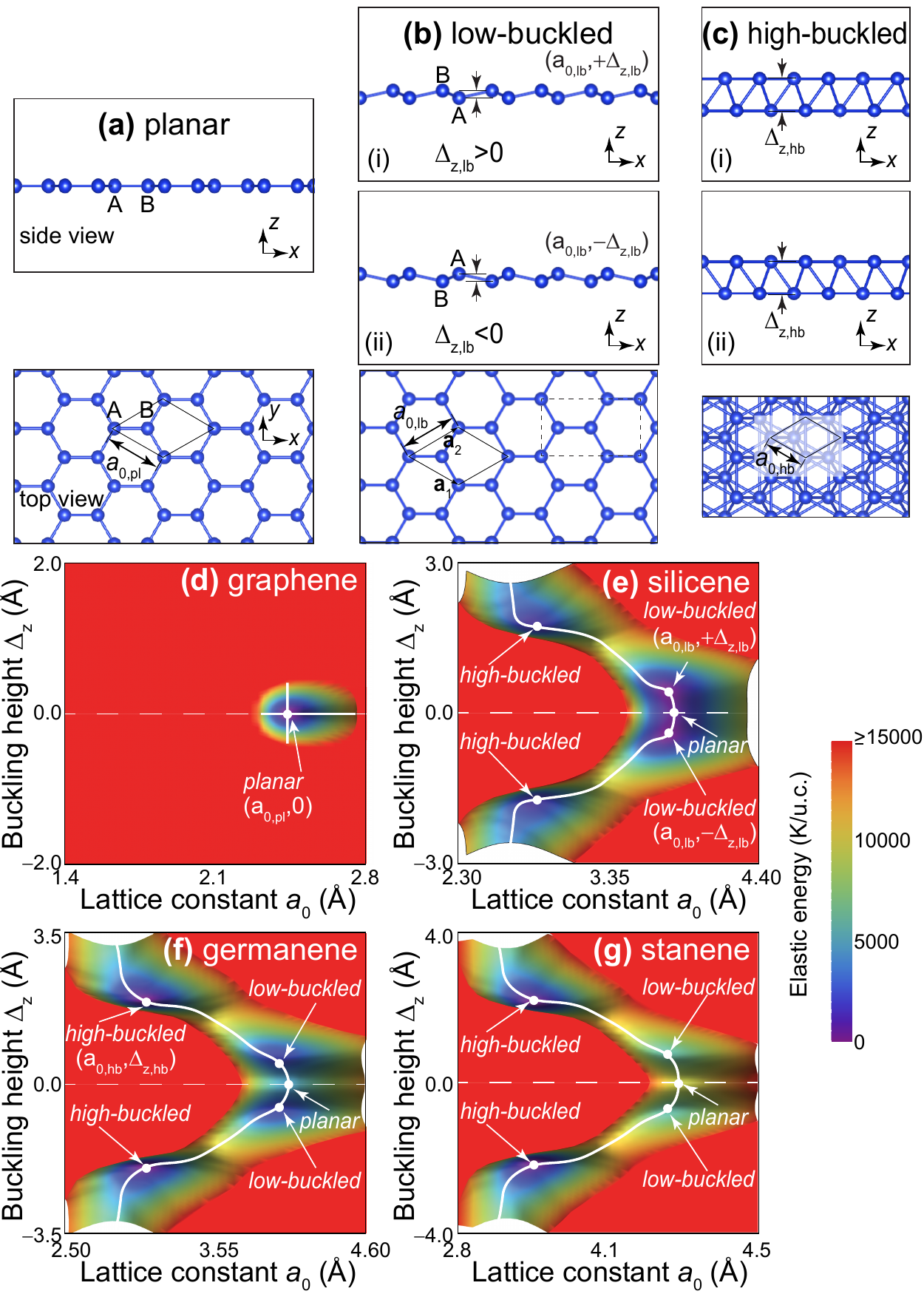}
\caption{Illustration of (a) planar and energy-degenerate (b) low-buckled, and (c) high-buckled two-dimensional lattices. Elastic energy landscape for (d) graphene, (e) silicene, (f) germanene, and (g) stanene. The horizontal axis on subplots (d-g) is $a_0$, while the vertical axis is $\Delta_z$. A path of steepest descent joining planar, low-buckled and high-buckled structures was overlaid on subplots (e) through (g) in white.}
\label{fig:1}
\end{figure}

Figure \ref{fig:1}(a) shows side and top views of a planar honeycomb structure. A primitive unit cell has been indicated by a diamond on the top view plot, with atoms on the $A-$ and $B-$sublattices indicated as well. The buckling height $\Delta_z$ is an order parameter, a {\em signed} measure of the relative displacement among $A-$ and $B-$sublattices along the $z-$direction, and it has a zero magnitude for planar structures: $\Delta_{z,pl}=0$ \AA. Fig.~\ref{fig:1}(b) depicts two degenerate three-fold coordinated, low-buckled structures, which are determined by two structural parameters: $a_{0,lb}$, and $\Delta_{z,lb}$, which is greater than zero on Fig.~\ref{fig:1}(b)(i), and smaller than zero on Fig.~\ref{fig:1}(b)(ii). We {\em define} $\Delta_z$ to be positive when the $B$ atoms sit above the $A-$atoms on the zero-temperature (initial MD frame) structure. A non-primitive, rectangular unit cell is shown in dashed lines on the top view as well. Fig.~\ref{fig:1}(c) shows two degenerate high-buckled, nine-fold coordinated two-dimensional structures \cite{RIVERO2015372}.

Figures \ref{fig:1}(d) through \ref{fig:1}(g) display elastic energy landscapes for graphene, silicene, germanene, and stanene, respectively. Energy landscapes were obtained by stretching the two in-plane primitive lattice vectors by the same amount--this is, by increasing the lattice constant $a_0$ only, without changing the $60^{\circ}$ angle among primitive lattice vectors $\mathbf{a}_1$ and $\mathbf{a}_2$ [see Fig.~\ref{fig:1}(b)]--while maintaining an identical distance and angular separation among the atoms belonging to opposite sublattices. Under those constraints, the lattice constant $a_0$ is shown as the horizontal axis, while $\Delta_z$ is represented by the vertical axis, and the false color serves to indicate the relative depth of global and/or local minima for each of these four two-dimensional materials. The landscapes on Figs.~\ref{fig:1}(d) through \ref{fig:1}(g) are similar to those published for other 2D materials with structural degeneracies before \cite{Mehboudi2016,prb2018,Tyler,Juan}.

\begin{figure}[tb]
\includegraphics[width=0.48\textwidth]{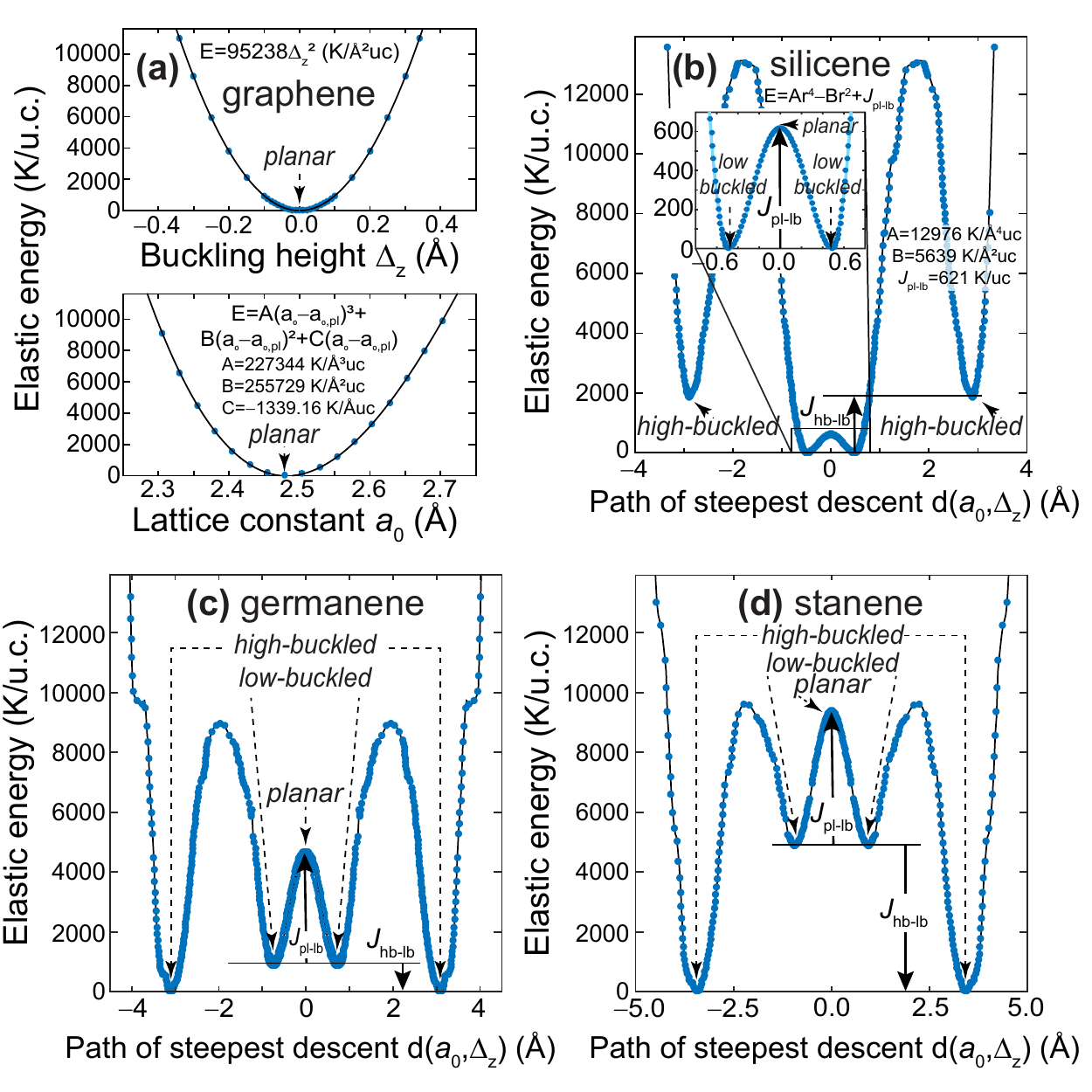}
\caption{One-dimensional cuts of the elastic energy landscapes along the steepest descent paths shown in white on Fig.~\ref{fig:1}. (a) Two cuts along orthogonal directions for graphene.
(b) Double-well elastic energy for silicene separating two low-buckled configurations with a barrier $J_{pl-lb}$ of 621 K/u.c.~(contrast against the profiles depicted in Refs.~\cite{prb2018,Tyler}, which show double-well potentials as well). High-buckled configurations require overcoming larger barriers. (c) and (d): Energy cuts for germanene and stanene, in which the high-buckled configurations are the lowest-energy ones, and the barrier $J_{hb-lb}$ to turn from a low-buckled configuration onto a high-buckled one becomes negative.
\label{fig:newf2}}
\end{figure}

Energy densities on Figs.~\ref{fig:1}(d-g) are reported in units of K/u.c., because a comparison of these values with critical temperatures $T_C$ obtained {\em via} molecular dynamics (obtained with the same {\em ab initio} numerical tool) gives a relation among the barriers $J$ obtained from the landscapes and $T_C$ of the order $J\leq T_C< 3J$ (see Refs.~\cite{Mehboudi2016,prb2018,PhysRevMaterials.3.124004} and \cite{review} for group-IV monochalcogenide monolayers; Refs.~\cite{seixas,Tyler,Alejandro} for SnO monolayers, and Ref.~\cite{Juan} for ferroelectric transition metal dichalcogenide bilayers). Areas on the landscape for which the energy exceeds 15,000 K/u.c.~are just shown in red. (This upper limit makes sense as the temperature at the sun’s surface is close to 6,000 K, and at that temperature materials would have long melted down). As seen on Figs.~\ref{fig:1}(e-g), the local minima can be traveled about within a ``trench'' on the energy landscape, and all the available local minima for buckled, hexagonal configurations are included in those plots. A one-dimensional path of steepest descent--similar to those drawn in Refs.~\cite{prb2018} and \cite{Tyler}--is shown in white as well.

To motivate the usefulness of the elastic energy landscape, and as seen on Fig.~\ref{fig:1}(d), graphene has a single (non-degenerate) structural ground state with a planar structure; a result that is expected. Silicene, on the other hand, displays {\em two degenerate global minima} upon a change of sign on $\Delta_{z,lb}$ in the low-buckled configuration [Fig.~\ref{fig:1}(e)]. These two distinguishable structures are related by a reflection with respect to the $xy$ plane, and {\em may be switchable} (making silicene a new {\em ferroelastic} two-dimensional material) by a deformation whereby the lattice constant increases slightly and the order parameter $\Delta_z$ becomes zero {\em on average}.

\begin{table}
\caption{Relaxed low-buckled and planar structural parameters at zero-temperature. Energy barriers in between low-buckled and planar (for which $\Delta_{z,pl}=0$) configurations are listed as well. See Fig.~\ref{fig:1} for the definition of structural parameters, and Fig.~\ref{fig:newf2} for a definition of the energy barrier $J_{pl-lb}$.
\label{table:lb_and_planar}}
\begin{tabular}{c c c c c}
\hline
\hline
Material & $a_{0,lb}$ (\AA) & $\Delta_{z,lb}$ (\AA) & $a_{0,pl}$ (\AA) & $J_{pl-lb}$ (\textrm{K/u.c.}) \\
\hline
graphene			& ---	&      0.000       & 2.475	         &  --- \\
silicene 			& 3.870	& $\pm$0.482       & 3.902	         &  621 \\
germanene 		    & 4.081	& $\pm$0.710       & 4.154	         & 3687 \\
stanene 			& 4.676	& $\pm$0.896       & 4.831	         & 4541	\\
\hline
\hline
\end{tabular}
\end{table}

\begin{table}
\caption{Relaxed high-buckled structural parameters at zero-temperature and energy barriers among low-buckled and high-buckled structures. Negative barriers indicate that the high-buckled structure is energetically preferred. See Fig.~\ref{fig:1} for the definition of structural parameters, and Fig.~\ref{fig:newf2} for a definition of the energy barrier $J_{hb-lb}$.
\label{table:hb}}
\begin{tabular}{c c c c}
\hline
\hline
Material & $a_{0,hb}$ (\AA) & $\Delta_{z,hb}$ (\AA) & $J_{hb-lb}$ (\textrm{K/u.c.}) \\
\hline
silicene 			& 2.688	& 2.143 &    1784 \\
germanene 		    & 3.022	& 2.246	& $-$1621 \\
stanene 			& 3.436	& 2.613 & $-$5448 \\
\hline
\hline
\end{tabular}
\end{table}

To start arguing for ferroelastic behavior on silicene--and for a possible two-dimensional phase transformation on this two-dimensional material--Fig.~\ref{fig:newf2} shows the elastic energy along the one-dimensional steepest descent paths $d=d(a_0,\Delta_z)$ depicted on Fig.~\ref{fig:1}(d) through Fig.~\ref{fig:1}(g). While graphene [Fig.~\ref{fig:newf2}(a)] has a single, non-degenerate energy minima on this landscape, silicene, germanene, and stanene display {\em four} local minima corresponding to low-buckled or high-buckled configurations.

As seen on Fig.~\ref{fig:newf2}(b), the energy region shown for low buckled silicene around $|d|\leq 0.8$ \AA{} displays an (anharmonic) double-well potential, which was fitted to:
\begin{equation}\label{eq:fit}
E(d)=12976\frac{K}{\text{\AA}^{4}\text{u.c.}}d^4 - 5639\frac{K}{\text{\AA}^2\text{u.c.}}d^2+621 \frac{K}{\text{u.c.}}
\end{equation}
The existence of two minima at $d=\pm 0.6$ \AA{} is a prelude of the bistable behavior of this 2D material \cite{prb2018,Tyler}, especially due to the low magnitude of the energy barrier $J_{pl-lb}$=621 K/u.c.~(see Table \ref{table:lb_and_planar}), which corresponds to $d=0$. Now, although germanene and stanene also have local minima at their low-buckled conformations, the barrier $J_{pl-lb}$ among them listed in Table \ref{table:lb_and_planar} is significantly higher (hindering the appearance of an average planar configuration), and high-buckled structures lie lower in energy. The differences in each subplot on Fig.~\ref{fig:newf2} underscore what should be different, distinct possible finite-temperature behaviors for each of those two-dimensional materials.

Table \ref{table:lb_and_planar} lists the optimal structural parameters at $(a_{0,lb},\pm \Delta_{z,lb})$; {\em i.e.}, at the (global or local) minima for low-buckled structures (with positive and negative signs on $\Delta_{z,lb}$ indicating the inherent two-fold structural degeneracy), the lattice constant $a_{0,pl}$ for the optimal (lowest energy) planar structure, and the energy $J_{pl-lb}$ (in units of Kelvin per primitive u.c.) needed to reach the planar structure from any of the degenerate low-buckled ones; see Fig.~\ref{fig:1}. Prior experience tells us that a critical temperature $T_c$ for a two-dimensional phase transformation onto another structure will be of the same order of magnitude than the barrier $J_{pl-lb}=E_{pl}-E_{lb}$, when $J_{pl-lb}$ is smaller than the melting point. $J_{pl-lb}$ is depicted in Fig.~\ref{fig:newf2}. According to Table \ref{table:lb_and_planar}, one may observe a transformation onto planar freestanding silicene at temperatures above 621 K.

Consistent with previous work \cite{bib:rivero}, the high-buckled phases--defined by the coordinates $(a_{0,hb},\pm\Delta_{z,hb})$ on Fig.~\ref{fig:1}--become preferable two-dimensional structure as the atomic mass of group-IV elements increases--see Figs.~\ref{fig:newf2}(c) and \ref{fig:newf2}(d). This statement is made quantitative by reporting the energy difference $J_{hb-lb}=E_{hb}-E_{lb}$ on Table \ref{table:hb}: As seen on Fig.~\ref{fig:newf2}, negative values of $J_{hb-lb}=E_{hb}-E_{lb}$ indicate a lower energy of the high-buckled 2D phase for both germanene and stanene. The consequence of the low-buckled phase turning metastable is that--starting on a low-buckled structure--germanene and stanene may melt, become amorphous, or turn into their high-buckled phase, rather than turning planar. Results of MD calculations, to be described next, shed light into the finite-temperature behavior of freestanding, low-buckled group-IV materials.

\section{Structural behavior of freestanding silicene, germanene, and stanene at finite temperature}\label{sec:MD}

We now contrast the information provided on Fig.~\ref{fig:1}, Fig.~\ref{fig:newf2},  Table \ref{table:lb_and_planar}, and Table \ref{table:hb}, with the finite-temperature behavior of low-buckled silicene, germanene, and stanene. This comparative study contains (i) instantaneous temperature, structural energy, supercell-averaged lattice constants for the primitive unit cell $\langle a_0 \rangle$, and supercell-averaged buckling height $\langle \Delta_z \rangle$ as a function of time at selected target temperatures; (ii) structural snapshots and pair-correlation functions; (iii) the evolution of $\langle a_0 \rangle$ and of $\langle \Delta_z \rangle$ as a function of temperature, to observe or discard a low-buckled-to-planar two-dimensional transition; and (iv) electronic structures. These four sets of information will be provided gradually for silicene (Sec.~\ref{sec:silicene}), germanene (Sec.~\ref{sec:germanene}), and stanene (Sec.~\ref{sec:stanene}) in what follows.

\subsection{Silicene's finite-temperature behavior}\label{sec:silicene}

\begin{figure}[tb]
\centering
\includegraphics[width=0.48\textwidth]{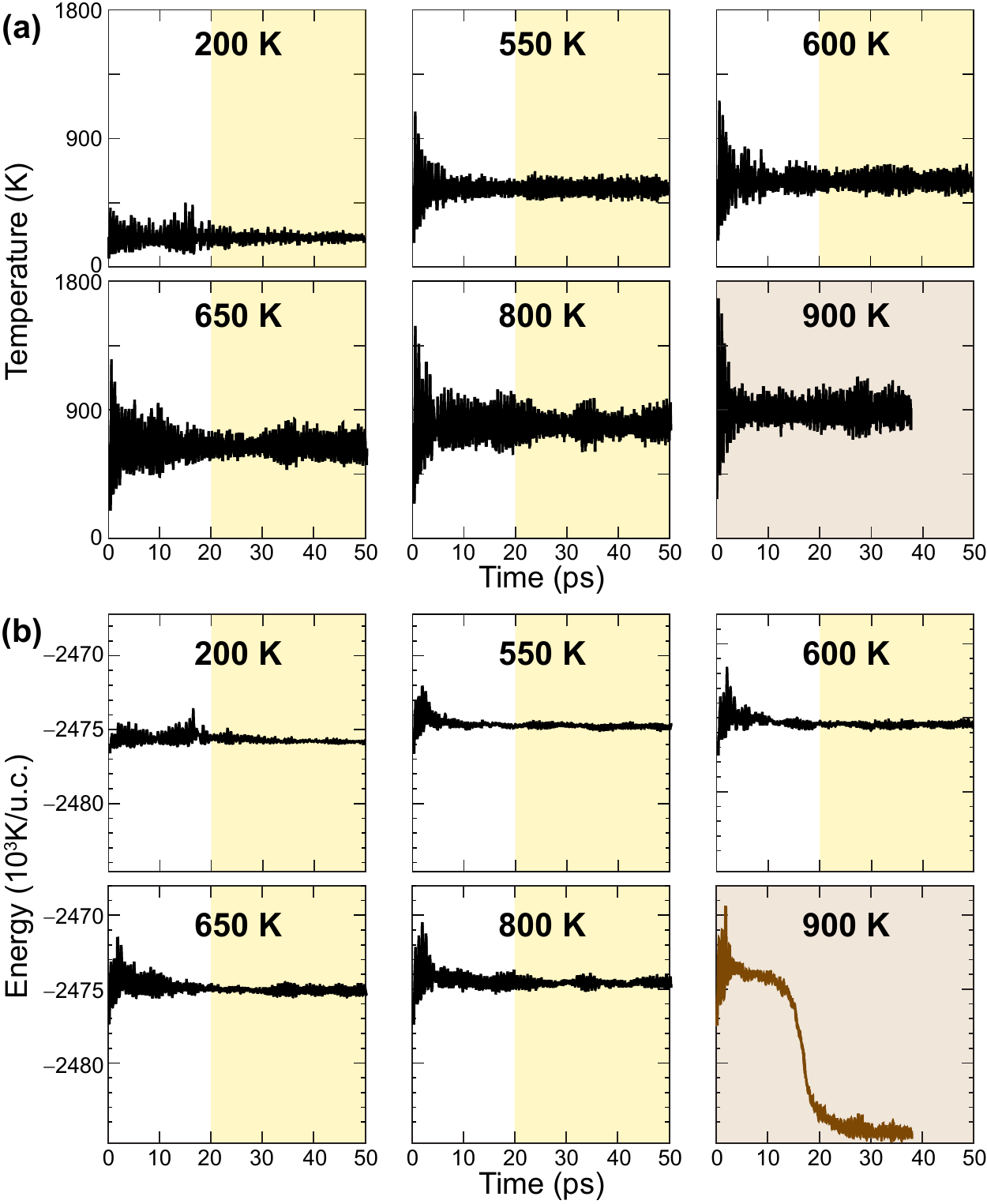}
\caption{Instantaneous (a) temperature and (b) structural energy for silicene at finite temperature; the 2D supercell structure was originally set in the low-buckled, zero-Kelvin configuration. Target temperatures are indicated in bold font. The yellow shading seen on subplots at 200, 550, 600, 650, and 800 K indicates the time interval that will be employed to calculate averages later on. At 900 K, the original time step resulted too large to properly converge the electronic density, and a smaller timestep was employed. The energy trace (drawn in brown) indicates a sudden decrease of energy, whose consequences on structure are highlighted on Fig.~\ref{fig:3}.\label{fig:2}}
\end{figure}

Fig.~\ref{fig:2}(a) displays the instantaneous temperature over a period of up to 50 ps for silicene supercells set at 200 K, 550 K, 600 K, 650 K, and 800 K, respectively. An additional plot at a target temperature of 900 K ends shortly before 40 ps for reasons that will be made clear in a moment. The takeaway point from Fig.~\ref{fig:2}(a) is that there are dramatic temperature fluctuations for the first 10 ps, and thermal equilibration sets in for longer times. Fig.~\ref{fig:2}(b) shows the time evolution of the instantaneous total energy per primitive u.c.~at 200 K, 550 K, 600 K, 650 K, 800 K, and 900 K: the total energy steadily increases with temperature for target temperatures up to 800 K. The marked decrease of the instantaneous total energy at 900 K on Fig.~\ref{fig:2}(b) is a sign of a sudden atomistic reconfiguration onto a more energetically favorable structure. Since atomistic forces on {\em ab initio} MD necessitate the convergence of the electronic density, we decreased the time step from 1.5 fs down to 0.5 fs at this higher temperature (since a smaller time step implies a smaller displacement, and an easier to converge electron density at each consecutive MD step), and collected the dynamics up to a shorter time as a result. Nevertheless, the relative flatness of the instantaneous energy above 30 ps on Fig.~\ref{fig:2}(b) at 900 K indicates that a further sudden drop of the total energy may not occur within the time window studied.

\begin{figure}
\includegraphics[width=0.48\textwidth]{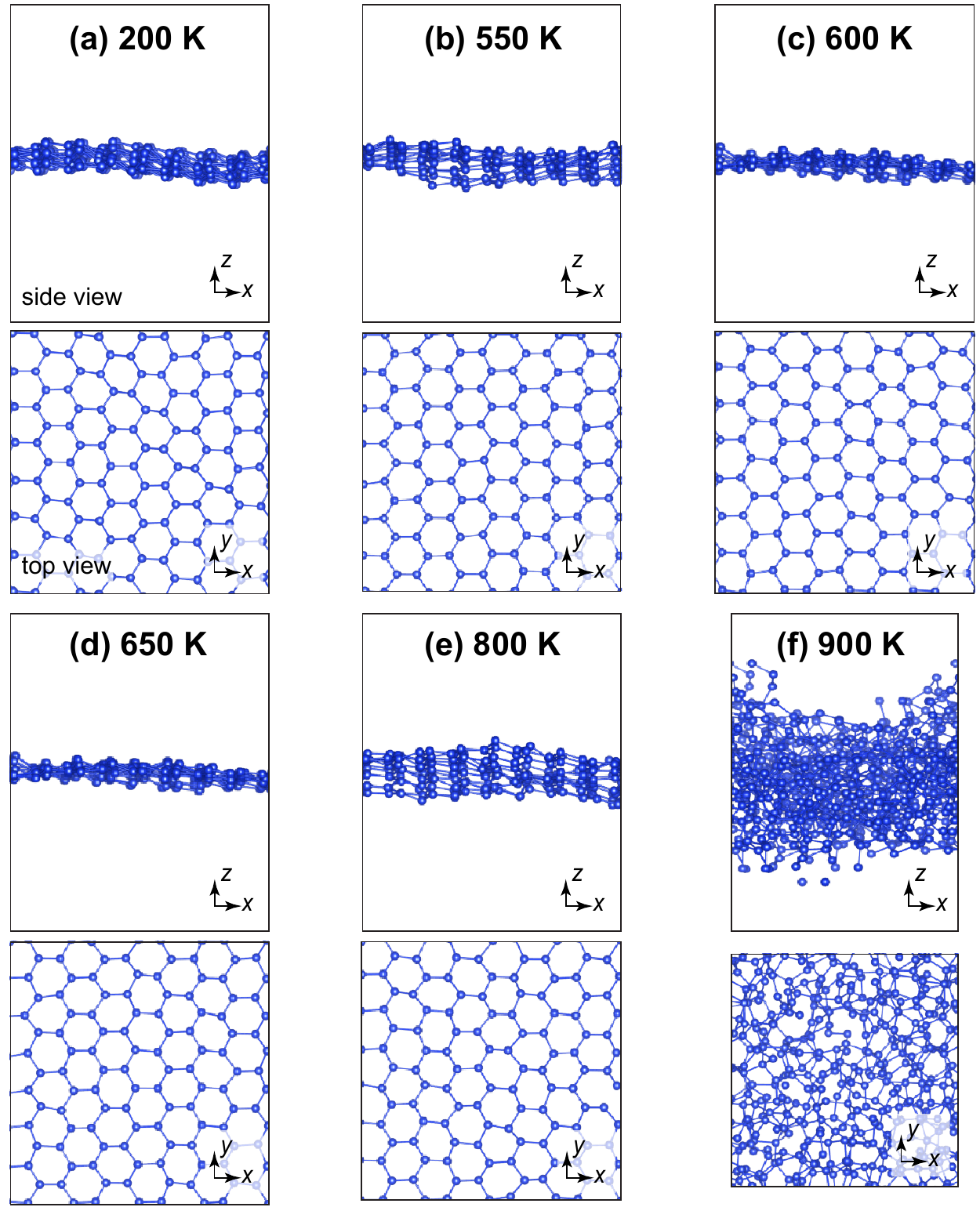}
\caption{Side and top snapshots at the last frame during simulations. The snapshots depict a $4\times 7$ portion of the ($7\times 12$) silicene supercell at 200, 550, 600, 650, 800, and 900 K target temperatures. Silicene turns amorphous at 900 K, explaining the sudden drop of energy on Fig.~\ref{fig:2}, and hinting at the inherent structural instability of 2D silicon.\label{fig:3}}
\end{figure}

\begin{figure}[tb]
\includegraphics[width=0.48\textwidth]{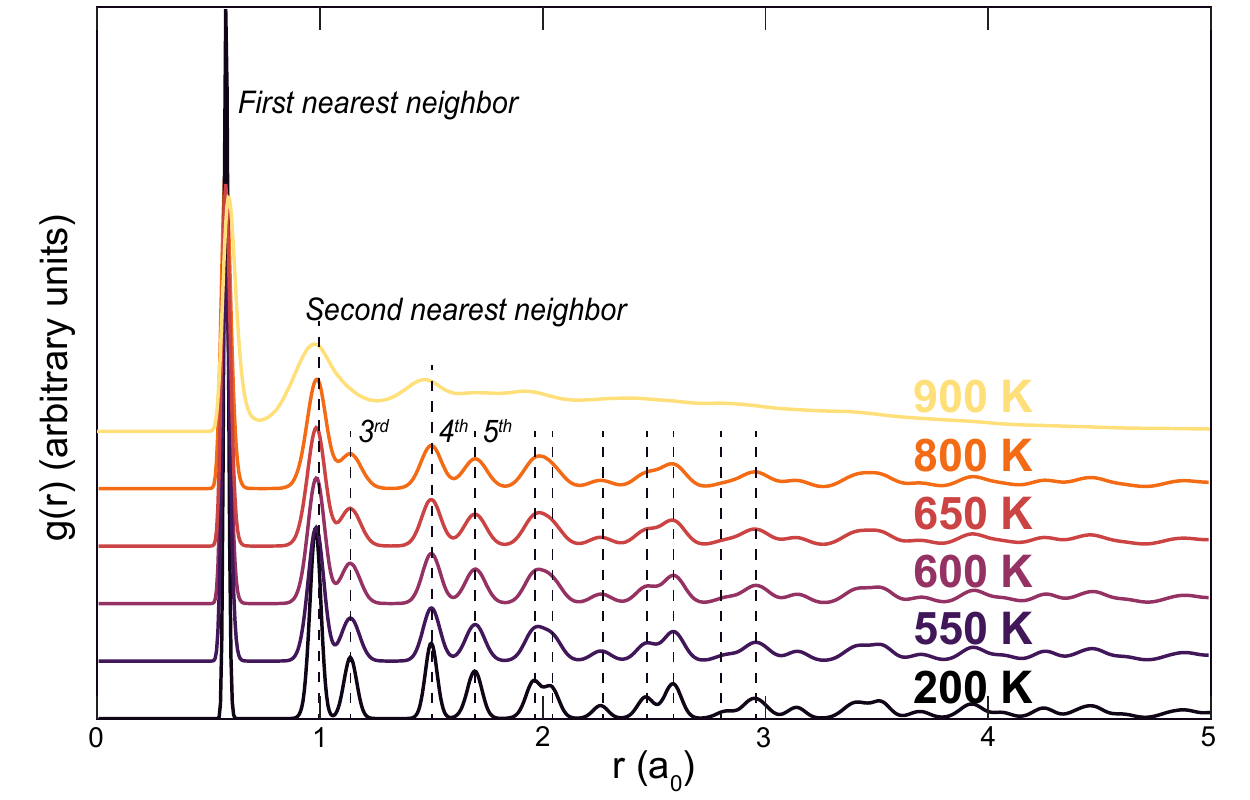}
\caption{Silicene pair correlation function $g(r)$ {\em versus} temperature. These are vertically offset from one another to better contrast the locations of neighboring atoms as temperature increases.  Well-resolved peaks for up to twelve nearest neighbors at 200, 550, 600, 650, and 800 K (dashed vertical lines) indicate that a hexagonal lattice is preserved. At 900 K, nevertheless, one observes only two well-defined peaks at the first and second nearest neighboring distances, which implies dimerization and the creation of an amorphous phase at that temperature.
 \label{fig:fig4}}
\end{figure}

To link the total energy to atomistic configurations, Fig.~\ref{fig:3} shows side and top view snapshots of the silicene supercell taken at the last simulated frame, at temperatures consistent with those shown on Fig.~\ref{fig:2}. The hexagonal lattice is visible at most temperatures, and the sudden drop of the total energy at 900 K on Fig.~\ref{fig:2}(b) is linked with a transition onto an amorphous phase [Fig.~\ref{fig:3}(f)], further confirmed by the pair-correlation functions $g(r)$ depicted on Fig.~\ref{fig:fig4}, which were vertically offset to better convey that the atomistic coordination remains almost unaltered for temperatures up to 800 K.

\begin{figure}[tb]
\centering
\includegraphics[width=0.48\textwidth]{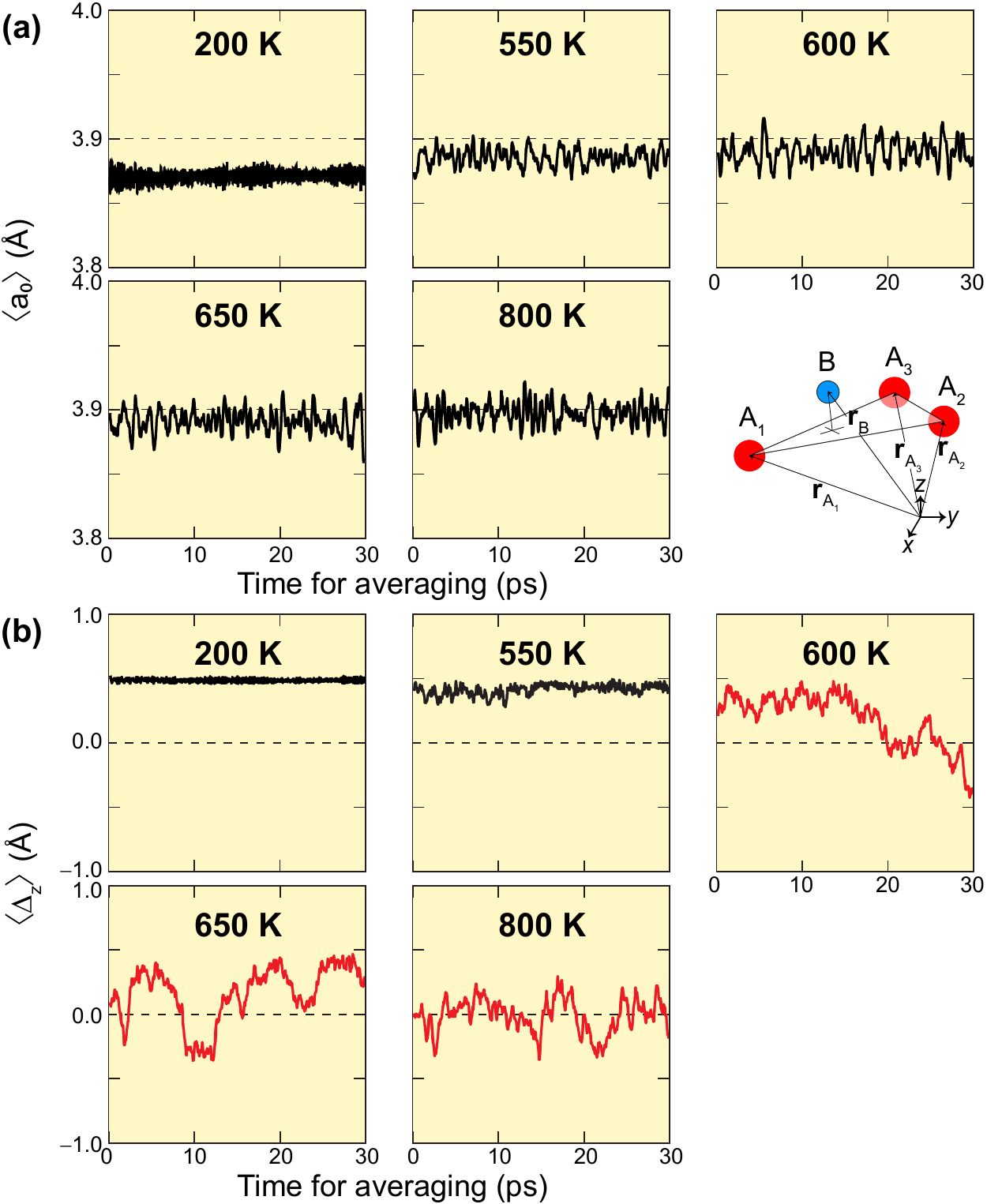}
\caption{(a) $a_0$ as a function of time and temperature. The inset at the lower right indicates a $B-$atom, and the three nearest $A$ atoms on the low-buckled configuration; see text for details. (b) Time evolution of order parameter $\Delta_z$ as a function of time at a given target temperature.  The yellow shading on all suplots stems from the fact that they were obtained past the first 20 ps; see Fig.~\ref{fig:2}. Red traces at 600 K, 650 K, and 800 K indicate that the average buckling height begins to take both positive and negative values, as the two wells seen for $|d|<0.6$ \AA{} on Fig.~\ref{fig:newf2}(b) begin to be both explored. The average ($\langle \cdot \rangle$) notation implies a supercell average of either $a_0$ or $\Delta_z$ at a fixed time during the MD evolution.\label{fig:fig5}}
\end{figure}

We argue that a two-dimensional phase transformation onto an {\em average planar structure} is taking place at around 600 K. Unveiling it requires paying close attention to the order parameter amounting to the relative height among atoms belonging to opposite sublattices {\em including sign}. Details follow.

\subsubsection{Relative height of the two atoms in the unit cell as an order parameter}

As a starting configuration, all supercells have all $B$ atoms above $A$ atoms at the onset of MD calculations [Fig.~\ref{fig:1}(b)(i)]. This means that every u.c.~has a positive value of $d$ on Fig.~\ref{fig:newf2}(b). This assertion is verified on Figs.~\ref{fig:1}(e-g), which show a value of the order parameter $\Delta_{z,lb}>0$. Correspondingly, negative values of $d$ on Fig.~\ref{fig:newf2} imply that {\em all} $B$ atoms on a crystal sit ``below'' nearest neighboring $A$ atoms. Identification of ``above'' and ``below'' is straightforward on an ideal crystal at zero temperature [in this particular situation, $\Delta_z=(\mathbf{r}_B-\mathbf{r}_A)\cdot \hat{z}$ gives away if the $B$ atom is above ($\Delta_z>0$), below ($\Delta_z<0$), or if the structure is planar ($\Delta_z=0$), where $\mathbf{r}_A$ ($\mathbf{r}_B$) is a position vector for any $A$ ($B$) atom, and $\hat{z}=(0,0,1)$], and it requires defining the projection of the position vector of $B$ atoms onto local signed normals set by the planes defined by the three closest $A$ atoms at finite temperature. ({\em Initial} structures for which a section contains $B$ atoms above $A$ atoms, while another section contains $A$ atoms above $B$ atoms, will have domain walls in which a pair of neighboring $A$ and $B$ atoms reside at the same height. Structures containing those domain walls were not considered here.)

While the supercells for MD calculations are {\em originally set with $\Delta_z>0$ at every primitive u.c.} (no domain behavior was considered), identification of unit cells for which $\Delta_z<0$ [$d>0$ on Fig.~\ref{fig:2}(b))] {\em at finite temperature} imply that $B$ atoms (originally above $A$ atoms) would have sufficient elastic energy to jump the double potential well on some u.c.s of the supercell and turn {\em below} $A$ atoms. This means that $B$ atoms on those u.c.s would be exploring the region $d<0$ on Fig.~\ref{fig:newf2}, setting the material at the onset of a two-dimensional structural transformation \cite{prb2018,Tyler}. Details follow.

\subsubsection{Transformation onto a planar two-dimensional silicene on average at a critical temperature}
A hexagonal lattice is created by two interconnected triangular lattices, and an atom belonging to the $B-$sublattice with coordinates $\mathbf{r}_B$ is always neighboring three atoms on the $A-$sublattice ($A_1$, $A_2$, and $A_3$) with coordinates $\mathbf{r}_{A_1}$, $\mathbf{r}_{A_2}$, and $\mathbf{r}_{A_3}$. As seen on the lower-right inset on Fig.~\ref{fig:fig5}(a), we choose the labels on the $A-$sublattice atoms to run counter-clockwise. Then, {\em local lattice vectors} $\mathbf{a}_1$ and $\mathbf{a}_2$ are defined as:
\begin{equation}
\mathbf{a}_1=\mathbf{r}_{A_2}-\mathbf{r}_{A_1},\text{ and }
\mathbf{a}_2=\mathbf{r}_{A_3}-\mathbf{r}_{A_1}.
\end{equation}
Local vectors $\mathbf{a}_1$ and $\mathbf{a}_2$ fluctuate at finite temperature. Panels on Fig.~\ref{fig:fig5}(a) display the lattice constant $\langle a_0 \rangle$ at finite temperature, which is obtained at each time step as the average of all possible local magnitudes of $|\mathbf{a}_1|$ and $|\mathbf{a}_2|$ within the $7\times 12$ supercell. Fluctuations of $\langle a_0\rangle$ are due to the fact that the containing walls are not fixed on the NPT ensemble, allowing for the low-buckled silicene supercell to slightly expand and compress as time goes on. The horizontal dashed line on Fig.~\ref{fig:fig5}(a) indicates the zero-temperature value of $a_{0,p}$ for planar silicene listed on Table \ref{table:lb_and_planar}, and one observes that $\langle a_0 \rangle$ raises onto $a_{0,p}$ as temperature increases.

Determining $\Delta_z$ requires specifying the {\em direction} of the local plane defined by the three atoms $A_1$, $A_2$, and $A_3$, and one gets one value of $\Delta_z$ per primitive u.c. We define the local plane's normal as:
\begin{equation}\label{Eq:1}
\hat{\mathbf{n}}=\frac{\mathbf{a}_1\times \mathbf{a}_2}{|\mathbf{a}_1\times \mathbf{a}_2|}.
\end{equation}
(Note that $\hat{\mathbf{n}}'=\frac{\mathbf{a}_2\times \mathbf{a}_1}{|\mathbf{a}_1\times \mathbf{a}_2|}$ is anti-parallel to $\hat{\mathbf{n}}$.) The (signed) local buckling height $\Delta_z$ is generalized for tilted honeycomb lattices at finite temperature as:
\begin{equation}\label{eq:updown1}
\Delta_z= (\mathbf{r}_{B}-\mathbf{r}_{A_1})\cdot\hat{\mathbf{n}}.
\end{equation}
See Appendix \ref{sec:appendixA} for details. Figure \ref{fig:fig5}(b) displays the average height $\langle \Delta_z \rangle$  over the supercell at a given temperature and time, revealing a situation in which $\langle \Delta_z \rangle$ suddenly fluctuates among positive and negative values for temperatures of 600 K or larger.

\begin{figure}[tb]
\includegraphics[width=0.48\textwidth]{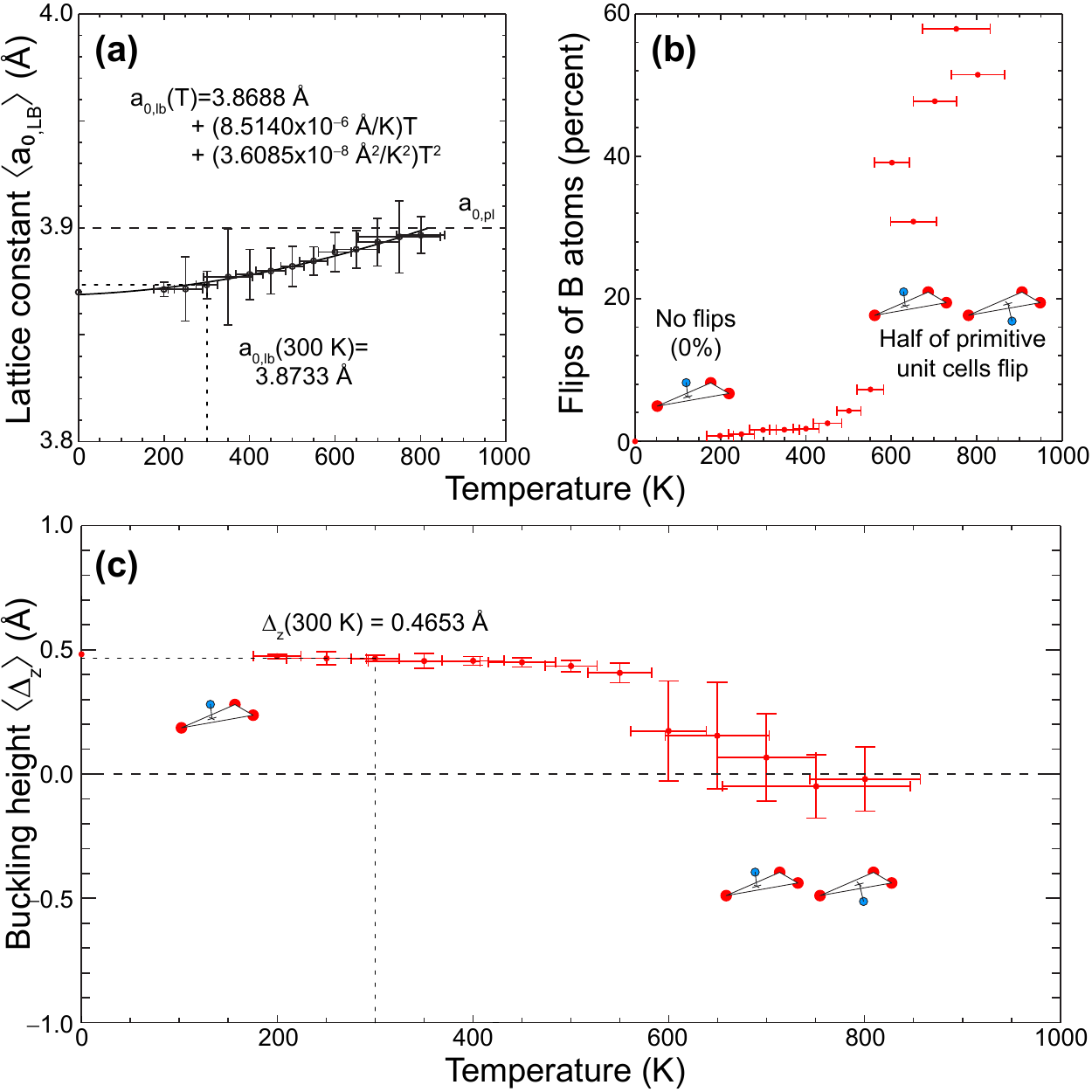}
\caption{(a) Thermal evolution of  $\langle a_0\rangle $ for silicene, initially set on its low-buckled structure. Average values and standard deviations from data presented on Fig.~\ref{fig:fig5}, and at additional intermediate temperatures are shown. Zero-temperature values for low-buckled and planar structures are taken from Table \ref{table:lb_and_planar}.
(b) Tracking of the percent of unit cells along for which the $B$ atom flips down below the nearest neighboring $A$ atoms: An equal distribution of ``above'' and ``below'' indicates that the two wells for $|d|<0.6$ \AA{} in Fig.~\ref{fig:newf2}(b) are equally sampled, and that the unit cell is planar on average. (c) A transition onto the planar structure is indicated by the  dramatic coalescence of the order parameter $\langle \Delta_z\rangle$ down to zero (with large fluctuations as well) for temperatures above $\sim$600 K, which is similar to the value of $J_{pl-lb}$ listed in Table \ref{table:lb_and_planar}. The structure turns amorphous at 900 K, and both $a_0$ and $\Delta_z$ are not properly defined then due to atomistic disorder; their values at that temperature are not shown for that reason.\label{fig:fig6}}
\end{figure}

Figure~\ref{fig:fig6}(a) shows $\langle a_0\rangle$ and its standard deviation, taken for the 20,000 frames captured within the last 30 ps (this is, after thermal equilibration; this time interval shown in yellow on Figs.~\ref{fig:2} and \ref{fig:fig5}). The dashed horizontal line on Fig.~\ref{fig:fig6}(a) once again indicates that the lattice constant of silicene steadily approaches the magnitude of the zero-temperature planar structure, $a_{0,p}$, as temperature increases. $a_{0_,lb}=3.8733$ \AA{} at ambient conditions.

As indicated earlier on, we originally set the silicene structure at the configuration indicated by a ``low-buckled'' label for $\Delta_z>0$ [Fig.~\ref{fig:1}(b)(i); point $(a_{0,lb},+\Delta_{z,lb})$ on Fig.~\ref{fig:1}(e); and $d\simeq 0.6$ \AA{} on Fig.~\ref{fig:2}(b)]. Furthermore, Table \ref{table:lb_and_planar} indicates that the planar structure can be created by adding and energy of 621 K per primitive unit cell. If that energy could be facilitated thermally, one would expect to reach a dynamical situation in which {\em both} low-buckled structural configurations [one in which $\Delta_z>0$ as in Fig.~\ref{fig:1}(b)(i); another in which $\Delta_z<0$ as in Fig.~\ref{fig:1}(b)(ii)] could be accessible with equal (50\%) probability. As seen on Fig.~\ref{fig:fig6}(b), such scenario takes place indeed. It reminisces of the 2D transformations discussed in Refs.~\cite{Mehboudi2016,Mehboudi2016b,editorial,prb2018} and \cite{Tyler} for other 2D materials in which a double-well potential with a small barrier is crossed at finite temperature, unleashing 2D transformations of the order-by-disorder type. Indeed, close observation to fluctuations of lattice parameters in Refs.~\cite{Mehboudi2016b} and \cite{Tyler} indicates that {\em square unit cells are, in fact, time averages of highly-fluctuating rectangular unit cells. Here, a planar structure is the average result of floppy vibrational modes enabled by a shallow degenerate wells on a Landau-type, similar elastic potential} [inset on Fig.~\ref{fig:newf2}(b)].

Fig.~\ref{fig:fig6}(c) shows the collapse of the order parameter $\langle \Delta_z \rangle$, further confirming the similarity of the phenomena observed in silicene and other phase-changing 2D materials \cite{Mehboudi2016,Mehboudi2016b,prb2018,Tyler,Juan} (see Table \ref{table:comparison} for a comparison). The collapse of $\langle \Delta_z \rangle$ past 600 K further confirms the use of the barrier $J_{pl-lb}$ among planar and low-buckled silicene to provide valuable intuition into the magnitude of the critical temperature. We are unaware of previous reports indicating such a two-dimensional structural transformation on silicene. At this point, we remind the reader that silicene is turning amorphous at 900 K (Fig.~\ref{fig:3}(f)), and that the barriers $J_{pl-lb}$ for both germanene and stanene exceed 3,500 K/u.c.~(Table \ref{table:lb_and_planar}): one should not expect a transition like the one we just documented for silicene on those other two-dimensional materials.

\begin{table}
\caption{Contrasting the existence of degenerate energies, order parameters, and change of coordination on novel 2D ferroelastic and ferroelectric materials against silicene's properties.\label{table:comparison}}
\begin{tabular}{c c c c}
\hline
\hline
2D Material                 & Degeneracy & Order par.       & Coord.~change? \\
\hline
SnO\cite{seixas,Tyler}      & Two-fold   & $\Delta \alpha$  & No                    \\
SnSe\cite{review}           & Two-fold   & $\Delta \alpha$  & 3-fold to 5-fold      \\
TMDC                        & Infinite   & Interlayer       & No                    \\
bilayers\cite{Wu,Juan}      &            & displacement     &                       \\
silicene                    & Two-fold   & $\Delta_z$       & No \\
\hline
\hline
\end{tabular}
\end{table}

\begin{figure}[tb]
\includegraphics[width=0.48\textwidth]{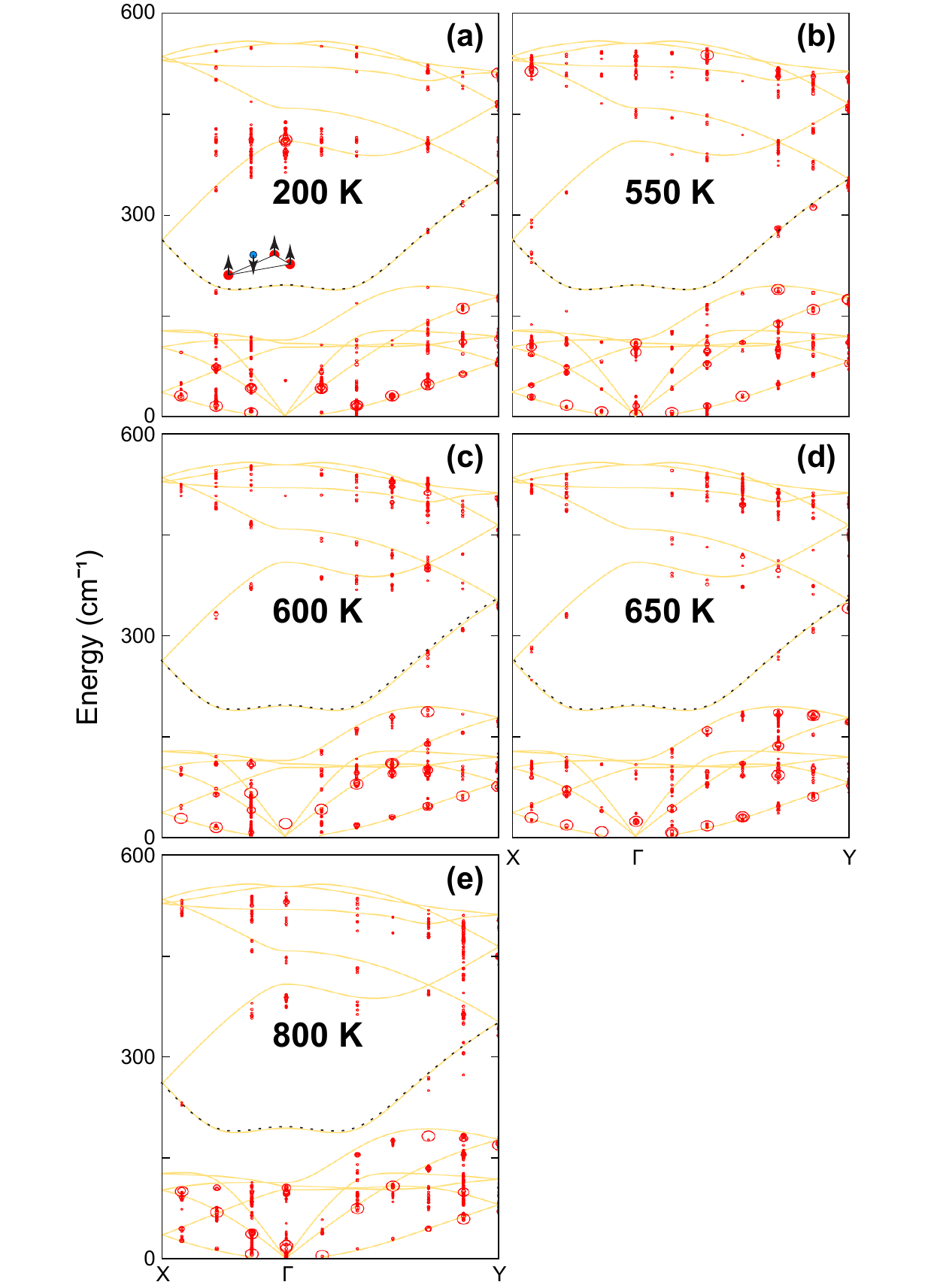}
\caption{Finite-temperature phonon modes for silicene obtained on a rectangular (non-primitive) unit cell containing four atoms. The diameter of the red circles indicates the relative intensity of a given power spectral line, and a collection of nearby circles tells the width (lifetime) of a given vibrational mode. The zero-temperature phonon spectra, obtained within the harmonic approximation, is shown by yellow curves. (Due to numerical precision, the lowest mode yields a minuscule, negligible, imaginary frequency of only 2.34 cm$^{-1}$ at $\Gamma$.) The out-of-plane optical mode leading to a planar configuration was highlighted by dotted lines--and by a diagram on subplot (a)--to better see that mode's softening at finite temperature.\label{fig:fig7}}
\end{figure}

\begin{figure}[tb]
\includegraphics[width=0.48\textwidth]{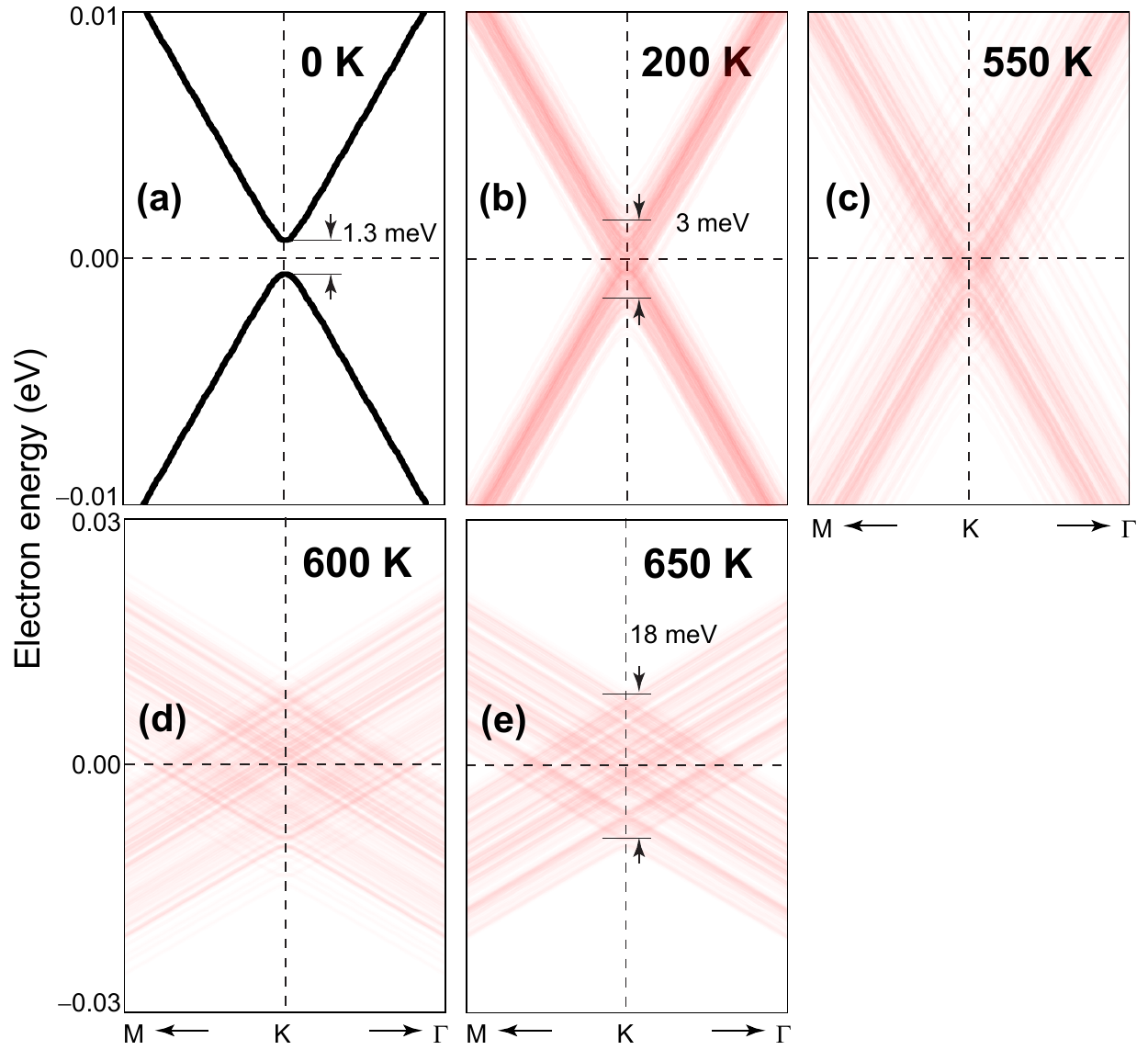}
\caption{Silicene electronic band structures around the $K-point$ at increasing temperature. For a given temperature, one hundred different band structures are overlapped to give a sense of broadening due to the atomistic vibrations. The 1.3 meV SOC-induced band gap of low-buckled silicene is obfuscated at every finite temperature due to the presence of electron and hole puddles. The band-broadening ranges from about 3 meV at 200 K, up to roughly 18 meV at 650 K, due to the disorder induced by the vertical flips above the transition temperature. At 600 and 650 K (for which the energy range was increased into $\pm0.03$ eV), the band gap is further reduced due to the advent of the planar phase; see Ref.~\cite{Huertas-Hernando} for details.\label{fig:fig8}}
\end{figure}

\begin{figure*}
\includegraphics[width=0.96\textwidth]{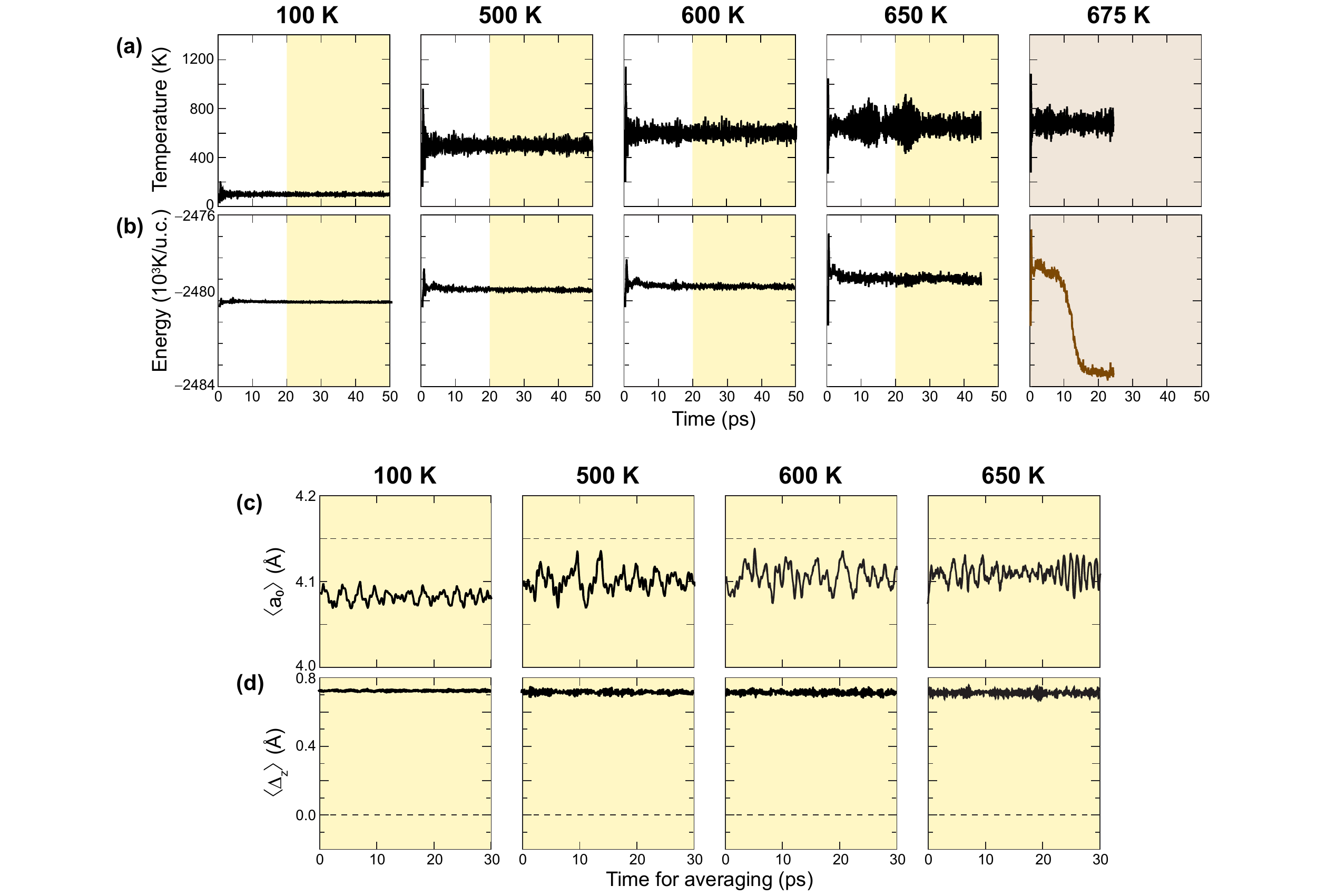}
\caption{Instantaneous (a) temperature, (b) structural energy, $\langle a_0 \rangle$, and (d) $\langle \Delta_z \rangle$ for germanene at finite temperature; the 2D supercell structure was originally set in the low-buckled, zero-Kelvin configuration. Target temperatures are indicated in bold font. The yellow shading seen on subplots at 100, 500, 600, and 650 K indicates the time interval that will be employed to calculate averages later on. At 675 K, the original time step resulted too large to properly converge the electronic density, and a smaller timestep was employed. The energy trace (drawn in brown) indicates a sudden decrease of energy, whose consequences on structure are discussed on Fig.~\ref{fig:fig10}.\label{fig:fig9}}
\end{figure*}

The existence of degenerate minima and a thermally reachable energy barrier necessarily guarantees anharmonic vibrational properties, and a strong electron-phonon coupling \cite{Thermoelectric}. Fig.~\ref{fig:fig7} shows both the vibrational spectra of low-buckled silicene as obtained within the harmonic approximation (yellow curves), and that obtained from the non-perturbative power spectrum of the velocity autocorrelation approach \cite{bib:koukaras} shown in (red) circles. The circles indicate the relative peak intensity, and the coalescence of multiple circles around a given zero-temperature phonon line is indicative of the lifetime of the vibrations obtained from MD. The choice of a rectangular cell on MD simulations was deliberate, as we wanted to simulate a near-square structure with equal dimensions to permit vibrations that could be homogeneous along the $x-$ and $y-$directions.

The softening of specific bands can be seen in multiple suplots of \ref{fig:fig7}. See, for instance, the dotted band corresponding to the out-of-plane optical mode, which softens in all plots (taking the average separation between the visible finite-temperature points and the continuous, zero-temperature lines on Fig.~\ref{fig:fig7}, it becomes smaller by 3.35 cm$^{-1}$ at 200 K, 4.17 cm$^{-1}$ at 550 K, 6.87 cm$^{-1}$ at 600 K, 5.40 cm$^{-1}$ at 650 K, and 9.26 cm$^{-1}$ at 800 K).

\subsubsection{Consequences for the SOC-induced bandgap}

The broadening of finite-temperature vibrational eigenmodes hinted at on Fig.~\ref{fig:fig7} should also broaden the electronic dispersion. To begin with, oscillations create strain and curvature, giving rise to electron-hole puddles arising from inhomogeneities of the local deformation potential on 2D materials  \cite{puddles0,puddles1,puddles2,puddles3,puddles4}. Those fluctuations of the deformation potential may be larger than silicene's zero-temperature 1.3 meV band gap, and it is a point we have not seen raised in the literature. In addition and according to Huertas-Hernando, Guinea, and Braatas, the intrinsic spin-orbit coupling can be greatly enhanced by buckling \cite{Huertas-Hernando}, and diminished in planar configurations.

Following an ergodic assumption--{\em i.e.}, that {\em spatial} averages give equivalent results to {\em time} averages--and using the method delineated in Appendix \ref{sec:defpot}, we present what perhaps is the most surprising result from the present work: that {\em atomistic vibrations obfuscate the 1.3 meV SOC-induced band gap of low-buckled silicene; this average closing of the SOC-induced bandgap occurs at a temperature as low as 200 K} [Fig.~\ref{fig:fig8}(b)]. In addition, SOC band gaps depend on curvature \cite{Huertas-Hernando}, and close on the planar configuration. This effect is clearly seen by comparing subplots \ref{fig:fig8}(a) with \ref{fig:fig8}(d) and \ref{fig:fig8}(e). The net fluctuations of the deformation potential are listed in Table \ref{table:Es} at Appendix \ref{sec:defpot}. That the minuscule band gap of silicene will be obscured by fluctuations of the deformation potential is an unquestionable result.

It is time to discuss how low-buckled germanene and low-buckled stanene behave at finite temperature in comparison. At this point, most methods have been introduced, which will give us the opportunity to mostly focus on different behaviors with respect to those seen in silicene.

\subsection{Germanene's finite-temperature behavior}\label{sec:germanene}

It is time to refer, once again, to Tables \ref{table:lb_and_planar} and \ref{table:hb}{, and to Figs.~\ref{fig:1}(f) and \ref{fig:newf2}(c)}. To begin with, the barrier to turn into a planar structure is much higher in germanene ($J_{pl-lb}=$3687 K/u.c.) than it is for silicene (621 K/u.c.): this should imply that one is {\em not} to observe a transition from a low-buckled onto a planar germanene phase. Furthermore, while the low-buckled phase of silicene was more stable than its high-buckled one \cite{bib:rivero}{, the negative sign of $J_{hb-lb}$ for germanene} in Table \ref{table:hb} says that a high-buckled phase should be more stable than the low-buckled structure set at the start of the MD evolution.

\begin{figure}
\includegraphics[width=0.48\textwidth]{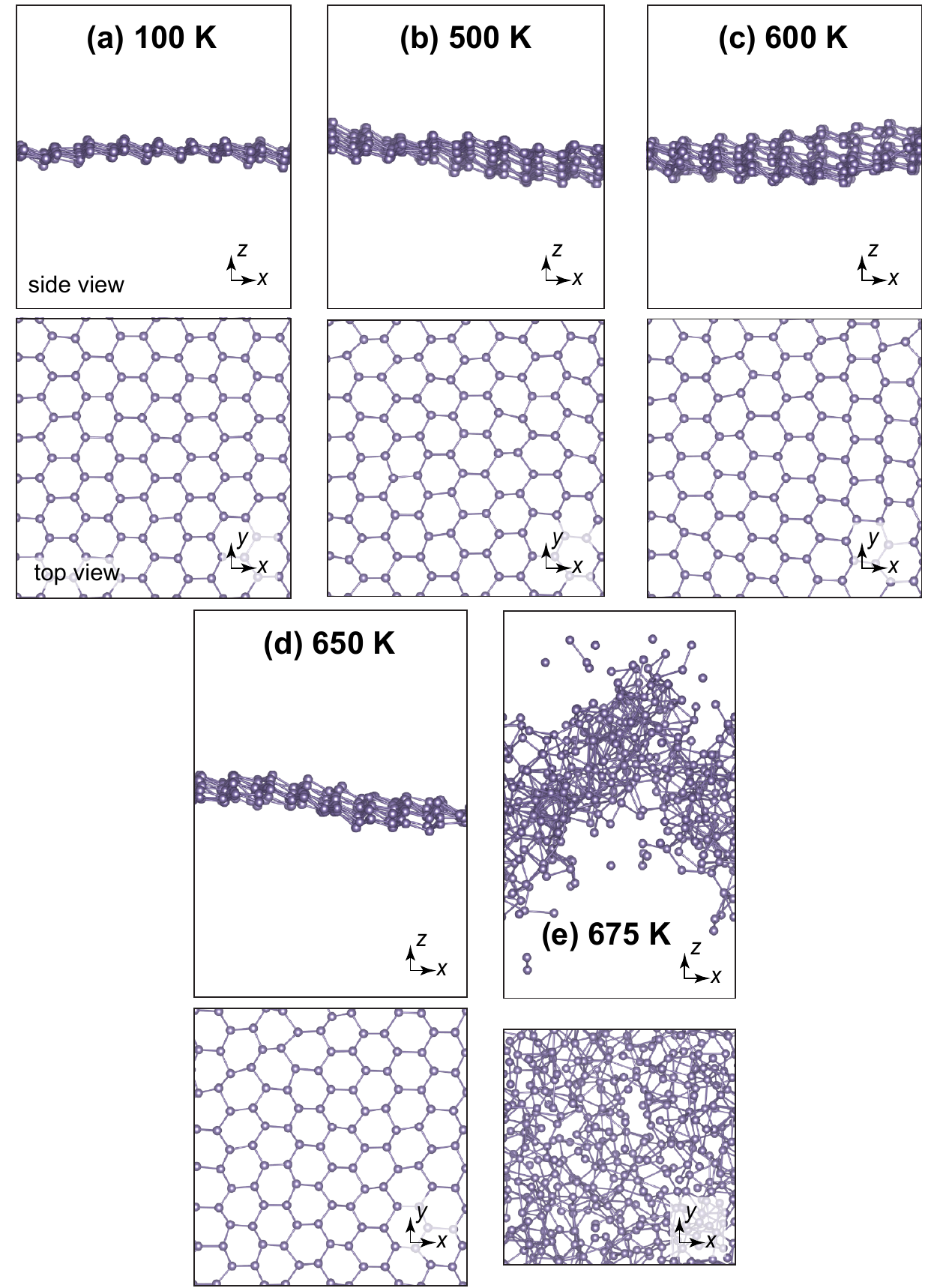}
\caption{Side and top snapshots at the last frame during germanene simulations. The snapshots depict a $4\times 7$ portion of the ($7\times 12$) supercell at 100, 500, 600, 650, and 675 K target temperatures. Germanene turns amorphous at 675 K, explaining the sudden drop of energy on Fig.~\ref{fig:fig9}(b), and showing the structural instability of 2D germanium.\label{fig:fig10}}
\end{figure}

\begin{figure}
\includegraphics[width=0.46\textwidth]{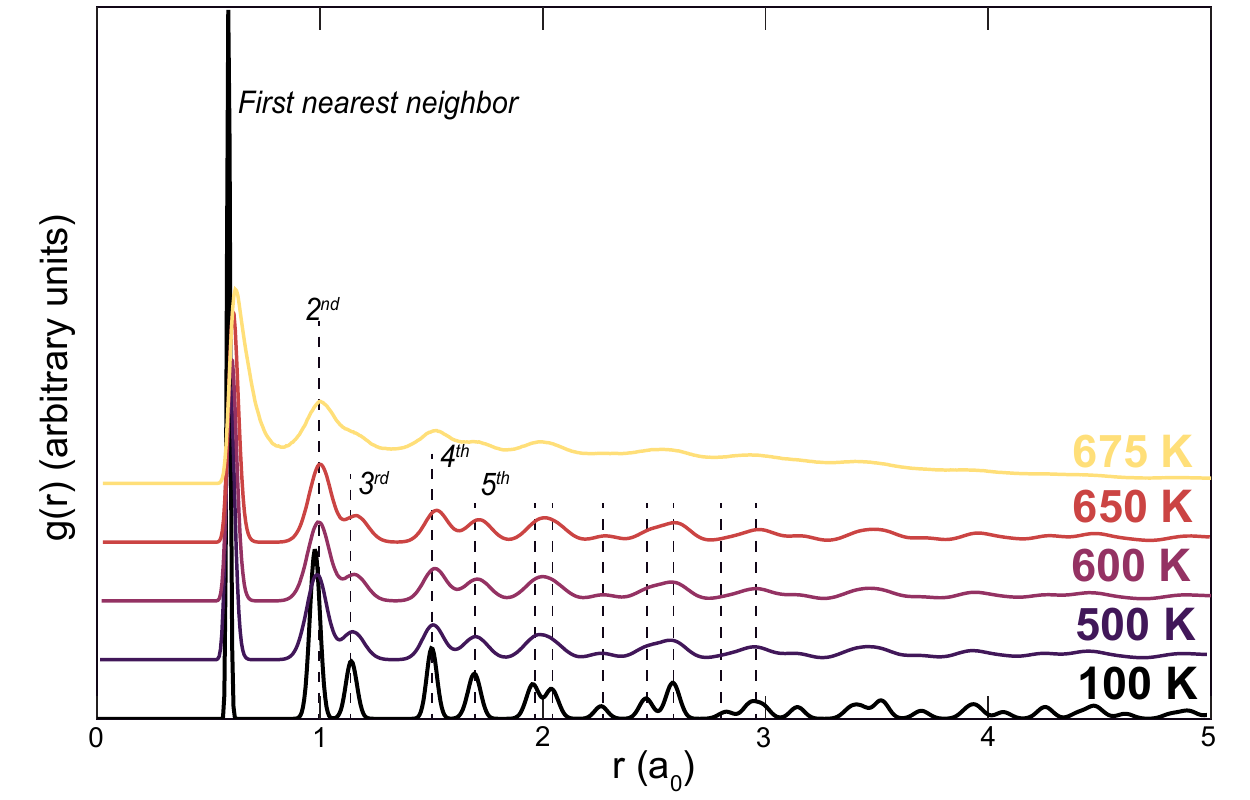}
\caption{Germanene pair correlation function $g(r)$ {\em versus} temperature. These are vertically offset from one another to better contrast the locations of neighboring atoms as temperature increases.  Well-resolved peaks for up to twelve nearest neighbors at 100, 500, 600, and 650 K (dashed vertical lines) indicate that a hexagonal lattice is preserved. At 675 K, an amorphous phase is revealed by the sudden loss of resolution of a number of nearest-atom peaks.\label{fig:fig11}}
\end{figure}

{Figure~\ref{fig:fig9}(a) shows the instantaneous temperature of germanene along the molecular dynamics simulation.} As seen on Fig.~\ref{fig:fig9}(b), the finite-temperature behavior of low-buckled germanene indicates a transition onto a lower-energy phase at 675 K, which is a lower temperature than that in which we observed the creation of amorphous silicon (900 K; see Fig.~\ref{fig:3}). Furthermore, for temperatures up to 650 K, {\em germanene never reaches a transition into a planar structure given that $J_{pl-lb}$=3687 K/u.c.} [see the distance to the horizontal dashed line on Fig.~\ref{fig:fig9}(c), and that $\langle \Delta_z \rangle$ never ever reduces its zero-magnitude value on Fig.~\ref{fig:fig9}(d)]. Verification of a transition of germanene onto an amorphous phase at 675 K is provided on Fig.~\ref{fig:fig10}(e), as well as on Fig.~\ref{fig:fig11}. {The fact that a transition onto an amorphous phase [see the sudden broadening of $g(r)$ in the 675 K trace on Fig.~\ref{fig:fig11}] takes place is related to the fact that the high-buckled phase is a ground state here, and that low-buckled germanene is a local (metastable) two-dimensional configuration; see Fig.~\ref{fig:newf2}(c).}

\begin{figure}[tb]
\includegraphics[width=0.48\textwidth]{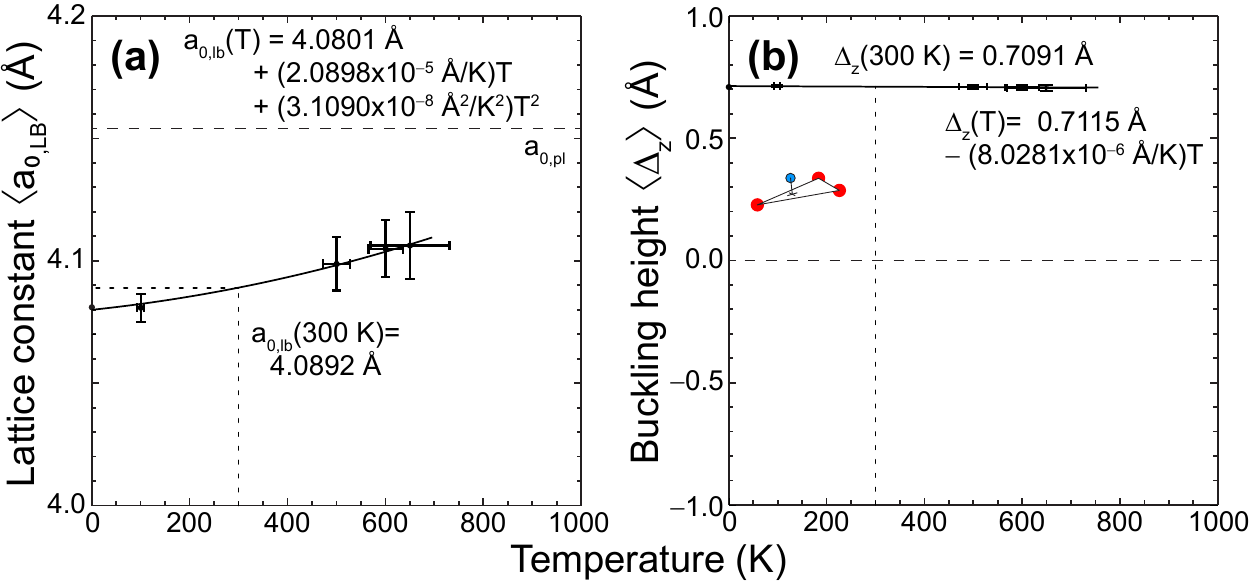}
\caption{Thermal evolution of (a) $\langle a_0\rangle $ and of (b) $\langle\Delta_z\rangle$ for germanene, initially set on its low-buckled structure. The data points shown are the average values and standard deviations from the data presented on Fig.~\ref{fig:fig5} and at additional intermediate temperatures. Zero-temperature values for low-buckled and planar structures are taken from Table \ref{table:lb_and_planar}, and room temperature values of $a_0$ and $\Delta_z$ are taken from the fits shown.\label{fig:newfig12}}
\end{figure}

{Figure \ref{fig:newfig12} demonstrates that thermal expansion alone is not sufficient for germanene to undergo a transition onto a planar structure: the expansion seen on Fig.~\ref{fig:newfig12}(a) is not sufficient to reach the magnitude $a_{0,pl}$ displayed as a horizontal line up to the temperature at which it melts. Nevertheless, $\Delta_z$ on Fig.~\ref{fig:newfig12}(b) remains almost unchanged up to the melting point. The lack of flips on germanene is explained by the tall elastic energy barrier $J_{pl-lb}$ observed on Fig.~\ref{fig:newf2}(c) and reported in Table \ref{table:lb_and_planar}. That it melts at a lower temperature than silicene may be underpinned by the fact that the low-buckled phase is a metastable one. Ambient values of $a_{0,pl}$ and of $\Delta_{z,pl}$ (4.0892 \AA{}, and 0.7091 \AA{}, respectively) can be found in Fig.~\ref{fig:newfig12} as well.}

\begin{figure}
\includegraphics[width=0.48\textwidth]{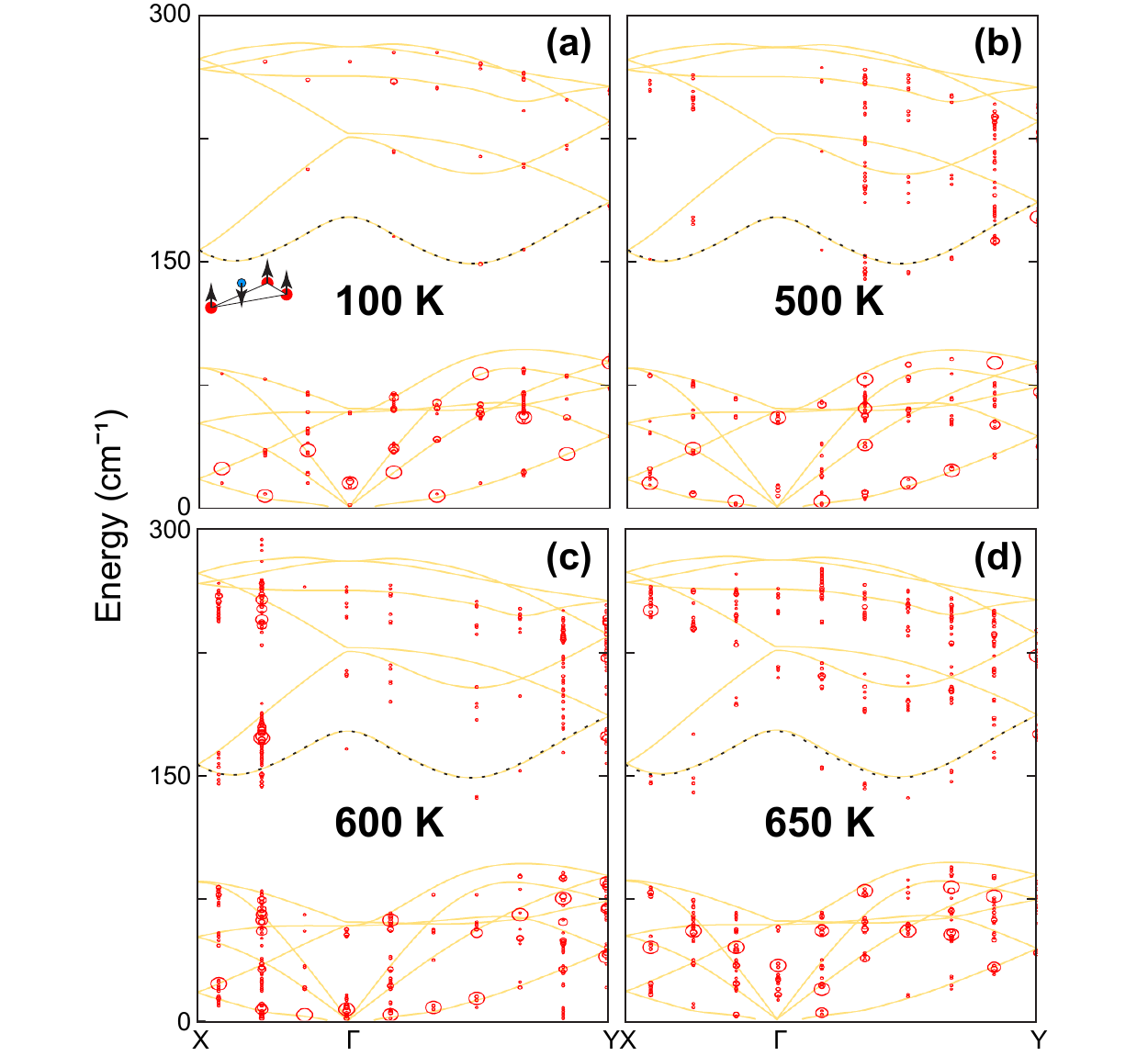}
\caption{Finite-temperature phonon modes for germanene obtained on a rectangular unit cell containing four atoms. The diameter of the red circles indicates the relative intensity of a given power spectral line, and a collection of nearby circles tells the width (lifetime) of a given vibrational mode. The zero-temperature phonon spectra, obtained within the harmonic approximation, is shown by yellow curves. {(Due to numerical precision, the lowest mode yields a minuscule, negligible, imaginary frequency of just 0.57 cm$^{-1}$ at $\Gamma$.) The out-of-plane optical mode leading to a planar configuration was highlighted by dotted lines--and by a diagram on subplot (a)--to better see that mode's softening at finite temperature.}\label{fig:fig13}}
\end{figure}

Figure \ref{fig:fig13} shows the vibrational properties of low-buckled germanene. Here, the softening of vibrational bands can be seen in a similar way than in silicene. {For example, the dotted band, corresponding to the optical out-of-plane mode softens as a function of temperature as follows: by -1.36 cm$^{-1}$ at 100 K, -6.19 cm$^{-1}$ at 500, -10.88 cm$^{-1}$ at 600, and by -10.72 cm$^{-1}$ at 650 K.}

\begin{figure}
\includegraphics[width=0.48\textwidth]{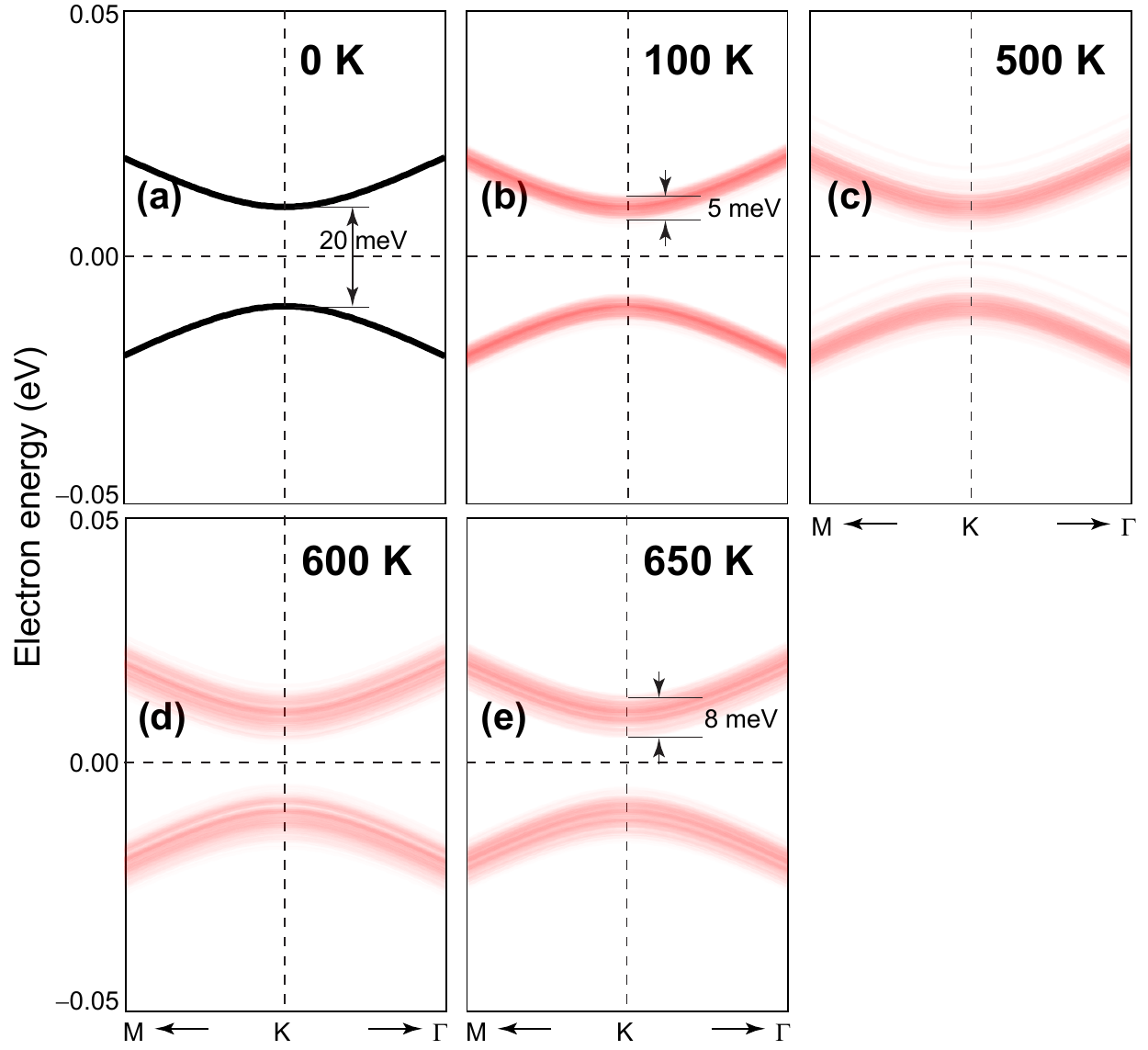}
\caption{Germanene electronic band structures around the $K-point$ at increasing temperature. For a given temperature, one hundred different band structures are overlapped to give a sense of broadening due to the atomistic vibrations. Unlike what was observed on Fig.~\ref{fig:fig8} for silicene, the SOC-induced band gap of low-buckled germanene is larger, and it persists up to the transition onto a disordered phase despite of atomistic vibrations. {The fact that $\Delta_z$ persists unchanged for germanene precludes the creation of additional fluctuations in the electronic band gap due to (above/below) flips among $A$ and $B$ atoms; see Ref.~\cite{Huertas-Hernando}.}\label{fig:fig14}}
\end{figure}

As for germanene's electronic properties, Fig.~\ref{fig:fig14} indicates that the SOC-induced {20 meV} band gap persists up to the melting point up to 650 K, despite of the observed phonon-induced broadening, {which is as large as 8 meV at 650 K and it is due to fluctuations of the deformation potential only. That silicene has a larger broadening at similar temperatures is a combined result of the deformation potential {\em and} the sudden changes in curvature; see Huertas-Hernando {\em et al.} \cite{Huertas-Hernando}}.

 This manuscript ends with a comparative discussion of low-buckled stanene at finite temperature.

\subsection{Stanene's finite-temperature behavior}\label{sec:stanene}
\begin{figure}
\includegraphics[width=0.48\textwidth]{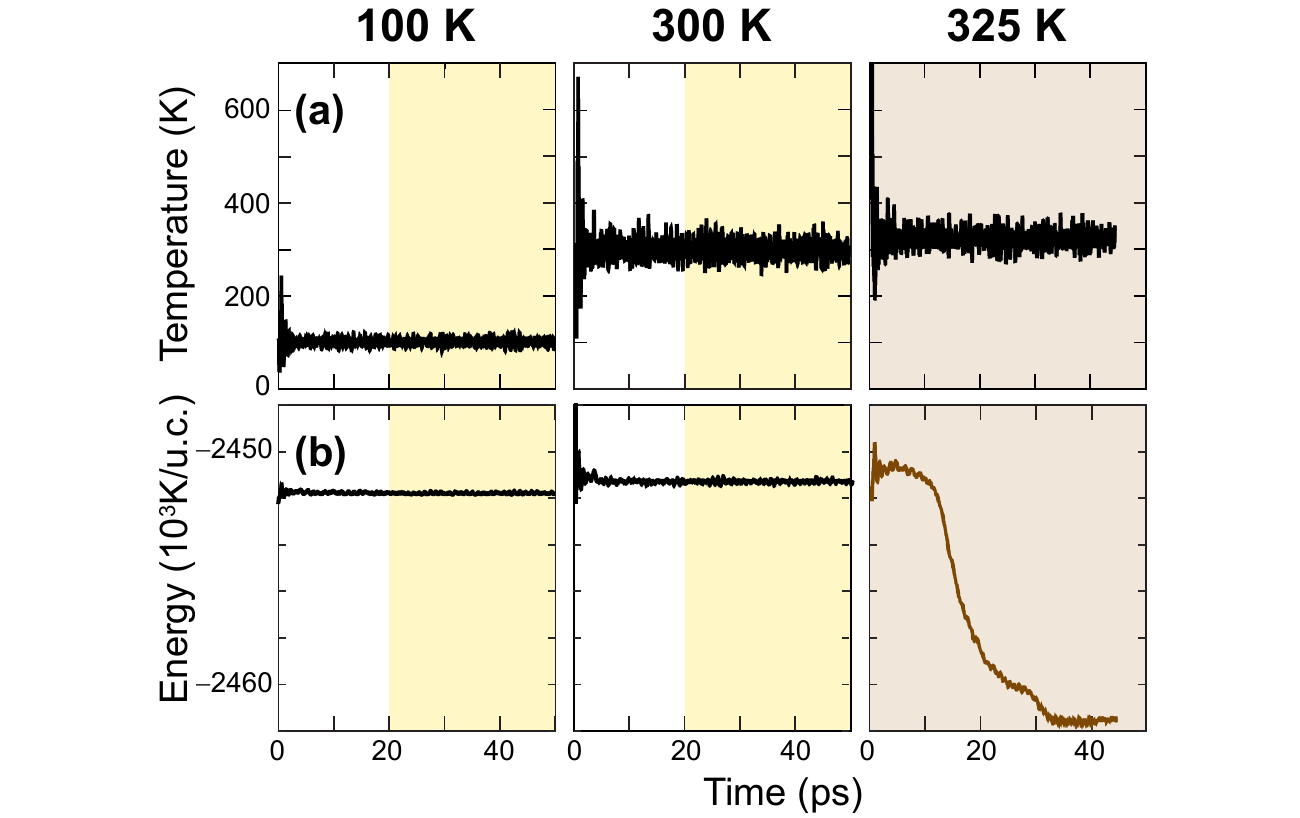}
\caption{Instantaneous (a) temperature, (b) structural energy, $\langle a_0 \rangle$, and (d) $\langle \Delta_z \rangle$ for stanene at finite temperature; the 2D supercell structure was originally set in the low-buckled, zero-Kelvin configuration. Target temperatures are indicated in bold font. The yellow shading seen on subplots at 100 and 300 K indicates the time interval employed to calculate time averages. At 325 K, the original time step resulted too large to properly converge the electronic density, and a smaller timestep was employed. The energy trace (drawn in brown) indicates a sudden decrease of energy, whose consequences on structure are seen on Fig.~\ref{fig:fig16}.\label{fig:fig15}}
\end{figure}

\begin{figure}
\includegraphics[width=0.48\textwidth]{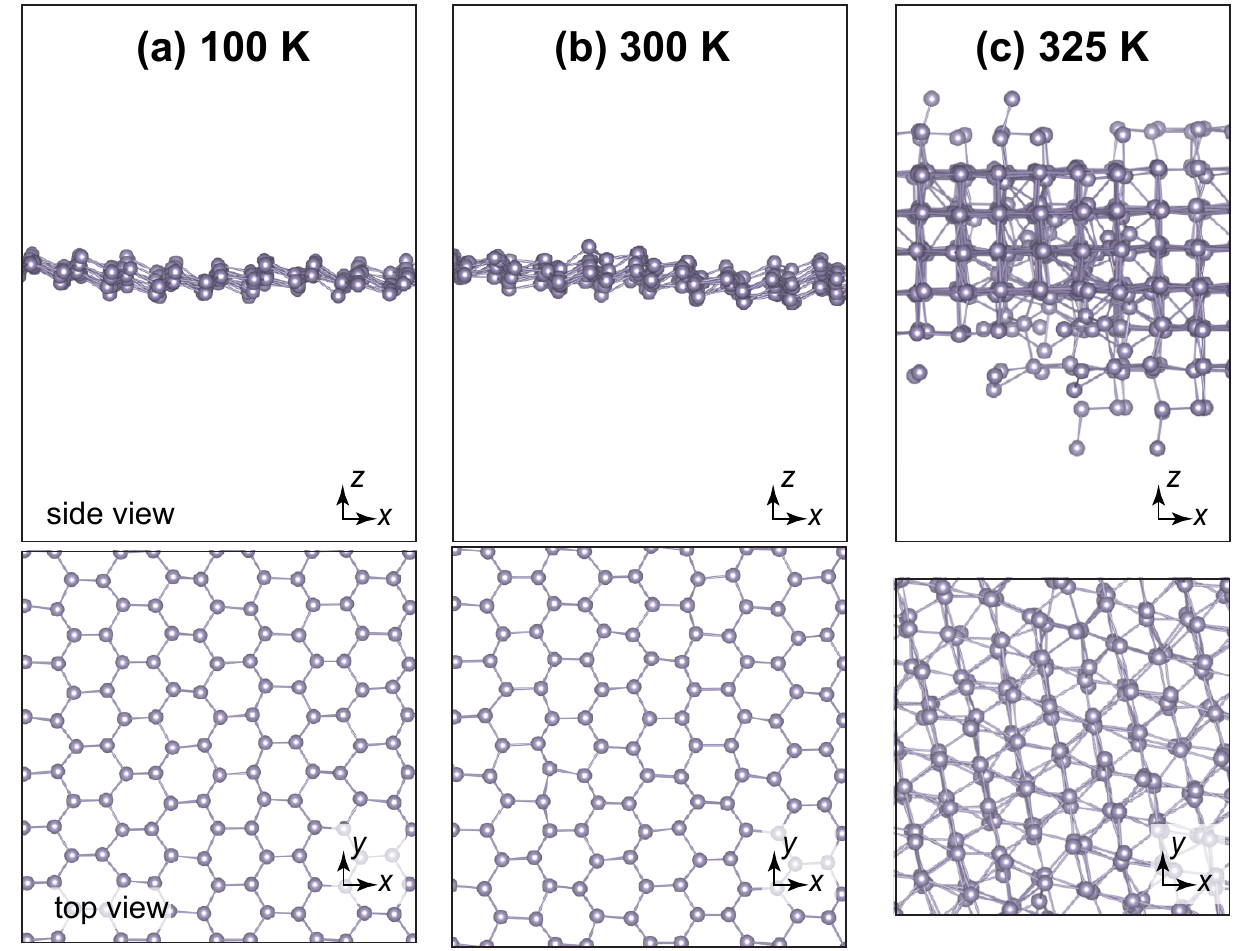}
\caption{Side and top snapshots at the last frame during stanene MD simulations. The snapshots depict a $4\times 7$ portion of the ($7\times 12$) supercell at 100, 300, and 325 K target temperatures. Stanene turns onto a bulk phase at 325 K, explaining the sudden drop of energy on Fig.~\ref{fig:fig15}(b), and showing the comparatively higher structural instability of 2D tin.\label{fig:fig16}}
\end{figure}

We refer to Tables \ref{table:lb_and_planar} and \ref{table:hb} one more time, {and to Figs.~\ref{fig:1}(g) and \ref{fig:newf2}(d),} to continue making the point that they contain crucial information that can be used to understand the finite-temperature behavior of low-buckled, freestanding, group-IV monoatomic two-dimensional materials.

The barrier to transition from a low-buckled onto a planar stanene configuration is a prohibitive 4541 K/u.c.~now. Furthermore, the high-buckled structure is now at a much lower,  5448 K/u.c.~below the low-buckled phase. This way, and as seen on Fig.~\ref{fig:fig15}, low-buckled stanene transitions onto another phase at just 325 K, without undergoing an intermediate planar phase (for which $\langle \Delta_z\rangle =0$) in the interim. {The structural plots at 325 K on Figure \ref{fig:fig16} confirm this observation, although a larger degree of order is observed when contrasted to silicene at 900 K [\ref{fig:3}(f)], or to germanene at 675 K [Fig.~\ref{fig:fig10}(e)].}

\begin{figure}
\includegraphics[width=0.48\textwidth]{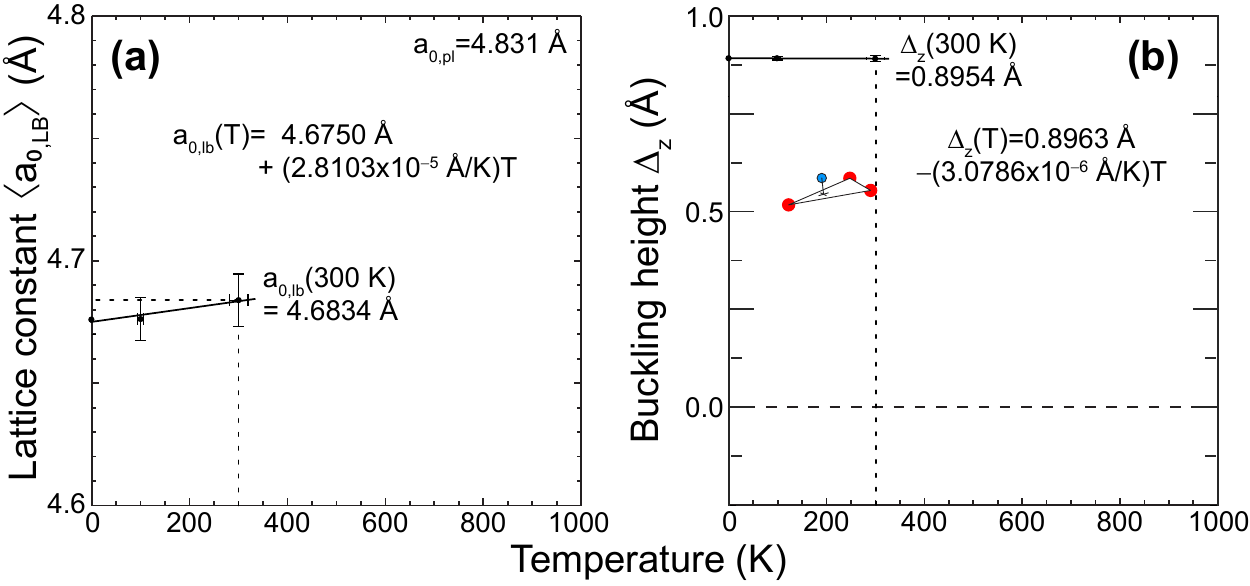}
\caption{Thermal evolution of (a) $\langle a_0\rangle $ and of (b) $\langle\Delta_z\rangle$ for stanene, initially set on its low-buckled structure. Zero-temperature values for low-buckled and planar structures are taken from Table \ref{table:lb_and_planar}. Stanene changed onto a thicker structure at 325 K.\label{fig:fig17}}
\end{figure}

\begin{figure}
\includegraphics[width=0.48\textwidth]{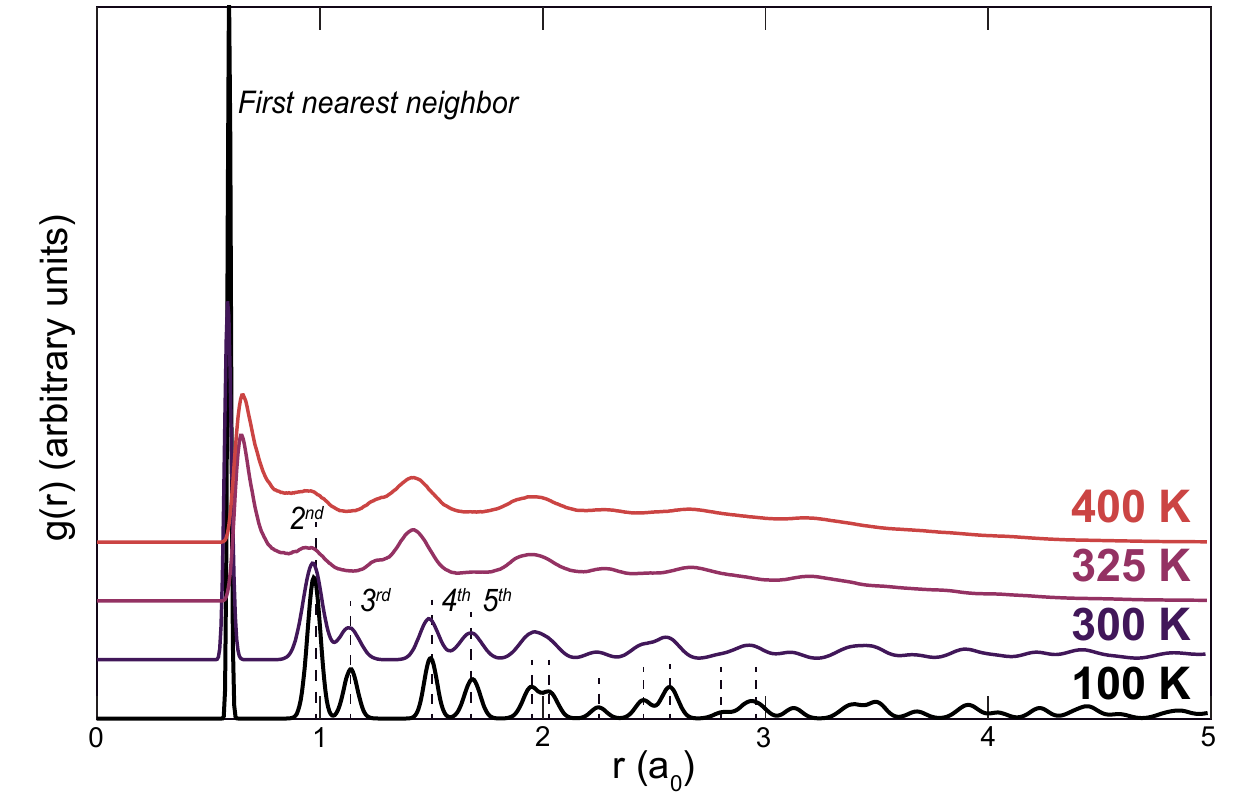}
\caption{Stanene pair correlation function $g(r)$ {\em versus} temperature. These are vertically offset from one another to better contrast the locations of neighboring atoms as temperature increases.  A transition onto a new ordered phase is revealed by well-defined peaks at shifted locations for the traces at 325 and 400 K.\label{fig:fig18}}
\end{figure}

Figure \ref{fig:fig17} contains average values for stanene's lattice constant and the buckling height, when started at the low-buckled configuration, and before it turns into the ordered, thicker phase observed on Fig.~\ref{fig:fig16}(c). {One observes the thermal expansion of $a_{0,lb}$ and an almost constant value of $\Delta_{z,lb}$, which remains stiff as in the case of germanene [Fig.~\ref{fig:newfig12}] due to the even taller elastic energy barrier $J_{pl-lb}=4541$ K/u.c.~[see Table \ref{table:lb_and_planar}], and unlike silicene [Fig.~\ref{fig:fig6}(c)] which was shown to undergo a transition whereby 50\% of the $B$ atoms are above $A$ atoms and 50\% lie below on average. The lattice constant and buckling height are equal to 4.6834 \AA{} and 0.8954 \AA{} at ambient conditions.}

The temperature-dependent pair correlation functions shown in Fig.~\ref{fig:fig18} differ from those shown on Figs.~\ref{fig:fig4} and \ref{fig:fig11}, in that {\em peaks develop at different distances}, as opposed to the gradual blurring that was observed for silicene and germanene. The displacements of the peaks that are observed in Fig.~\ref{fig:fig18} support the fact that a new, thicker and (hexagonal-closed-packed) bulk-like phase is being generated on tin at 325 K. This fact is further confirmed by showing a pair correlation function at a higher temperature of 400 K.

\begin{figure}
\includegraphics[width=0.48\textwidth]{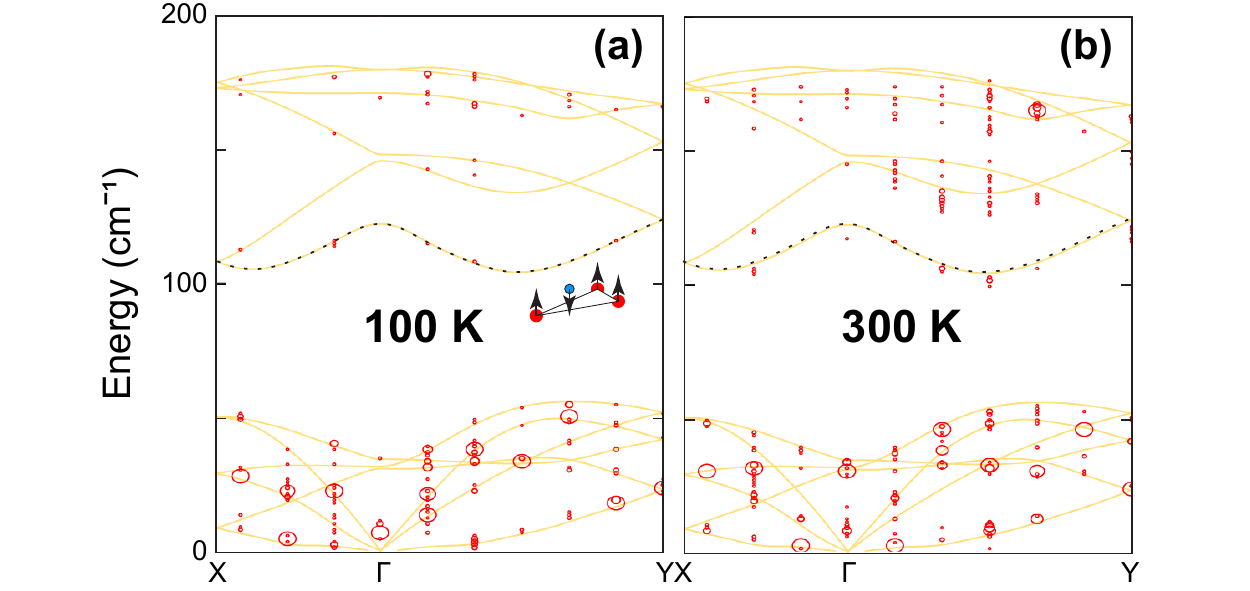}
\caption{Finite-temperature phonon modes for stanene obtained on a rectangular unit cell containing four atoms. The diameter of the red circles indicates the relative intensity of a given power spectral line, and a collection of nearby circles tells the width (lifetime) of a given vibrational mode. The zero-temperature phonon spectra, obtained within the harmonic approximation, is shown by yellow curves. {(Due to numerical precision, the lowest mode yields a minuscule, negligible, imaginary frequency of just 0.29 cm$^{-1}$ at $\Gamma$.) The out-of-plane optical mode leading to a planar configuration was highlighted by dotted lines--and by a diagram on subplot (a)--to better see that mode's softening at finite temperature.}\label{fig:fig19}}
\end{figure}

\begin{figure}
\includegraphics[width=0.48\textwidth]{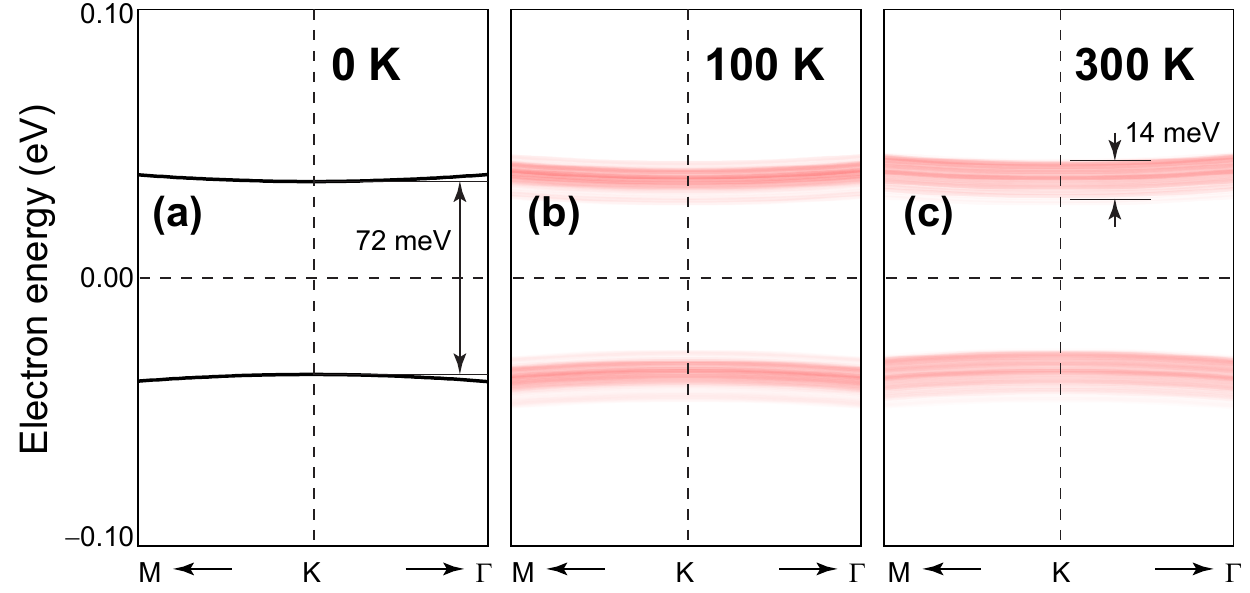}
\caption{Stanene electronic band structures around the $K-point$ at increasing temperature. For a given temperature, one hundred different band structures are overlapped to give a sense of broadening due to the atomistic vibrations. Similar to germanene, the SOC-induced band gap of low-buckled stanene persists up to the transition onto a bulk-like phase. {The fact that $\Delta_z$ persists unchanged for stanene precludes the creation of additional fluctuations in the electronic band gap due to (above/below) flips among $A$ and $B$ atoms; see Ref.~\cite{Huertas-Hernando}.}\label{fig:fig20}}
\end{figure}

The last point of discussion concerns the vibrational and electronic properties of low-buckled stanene prior to its structural transition onto a thicker structure. Figure \ref{fig:fig19} shows that the vibrational modes are consistent with those of the low-buckled structure up to room temperature, {and the mode highlighted by dotted lines softens by 0.56 cm$^{-1}$ at 300 K, and by 3.01 cm$^{-1}$ at 300 K.} Figure \ref{fig:fig20} shows that the {72 meV} SOC-induced bandgap remains robust despite of the electron-phonon coupling {because of the lack of vertical flips [see Fig.~\ref{fig:fig17}(b)]}.

 This way, we have provided a crucial assessment of 2D topological insulators at finite temperature, and provided guidance as to possible structural transformations that may affect intended topological properties of those materials on their freestanding form. The observations made using MD on this work support and complement those made earlier on \cite{bib:rivero}.

\section{Conclusion}\label{sec:conclusion}
Exhaustive molecular dynamics simulations and zero-temperature calculations were employed to understand the finite-temperature behavior of silicene, germanene, and stanene, initially set into their low-buckled atomistic configuration. The structural degeneracy (positive and negative buckling) suggested a possibility of ferroelastic behavior on these materials, anharmonic vibrational properties, and phase transitions.

Ferroelastic behavior, and a transition onto an average planar 2D structure was indeed observed in freestanding low-buckled silicene above 600 K. Numerical evidence shows that it melts at 900 K. We also found that the 1.3 meV SOC-induced bandgap of silicene is smaller in magnitude than fluctuations of the local deformation potential, which will hinder observation of said bandgap at temperatures as small as 200 K.

The high-buckled two-dimensional structure is preferable for germanene and stanene. Germanene reaches its melting phase transformation at 675 K, and it does not transition onto a planar phase at intermediate temperatures. The SOC-induced bandgap is robust against thermal oscillations.

Stanene reconfigures itself above room temperature (325 K) into a bulk HCP crystal, as is expected of heavier metallic materials. We did not observe ferroelastic behavior for germanene nor stanene.

Those results could be useful to link these material's expected atomistic and electronic structure at finite temperature (especially considering operation temperatures on actual devices), and gives information about expected phase transitions onto planar 2D structures (silicene), disordered ones (silicene and germanene), or bulk (stanene) phases.


\section{Acknowledgments}
We thank Dr.~Pradeep Kumar for insightful conversations, and Shiva P.~Poudel for technical assistance. Work funded by the U.S.~Department of Energy (Award DE-SC0022120).

\appendix
\section{Determination of the local value of $\Delta_z$}\label{sec:appendixA}
Taking the normal to the plane defined by atoms $A_1$, $A_2$, and $A_3$ as given in Eqn.~\eqref{Eq:1}, the equation of the plane spanned by these three points is given by:
\begin{equation}\label{eq:plane}
\left[ \mathbf{r} -(\mathbf{r}_{A_1}\cdot \hat{\mathbf{n}}) \hat{\mathbf{n}} \right]\cdot \hat{\mathbf{n}}=0.
\end{equation}
In Eqn.~\eqref{eq:plane}, $\mathbf{r}=(x,y,z)$ is a coordinate within the plane. Indeed, Eqn.~\eqref{eq:plane} can be used to solve for, say, $z$ as a function of $x$ and $y$; $x$ and $y$ then become the two independent variables defining the plane.

We next need to determine whether point $B$ lies ``{\em below}'' or ``{\em above}'' the plane defined by $A_1$, $A_2$, and $A_3$. Quotation marks are added because, as seen on the side views of Figs.~\ref{fig:3}, \ref{fig:fig10}, and \ref{fig:fig16}, the planes defined by any three neighboring atoms on the $A-$sublattice tilt with respect to the $xy$ plane one had at zero temperature, requiring additional clarification on the meaning of an atom being ``up'' or ``down'' a plane. We will use the normal vector $\hat{\mathbf{n}}$ to set a formal definition of an atom being up or down a plane next.

To do so, we need to find a specific point within the plane, which we will call $\mathbf{r}_0$, such that the distance between point $\mathbf{r}_B$ and $\mathbf{r}_0$ takes its minimum value. This condition occurs when the vector $\mathbf{r}_B-\mathbf{r}_0$ is parallel or antiparallel to $\hat{\mathbf{n}}$.

Now, a parametrization of a straight line containing the vector $\hat{\mathbf{n}}$ is given by $\lambda \hat{\mathbf{n}}$, with $\lambda$ a real number. This line does not necessarily pass through point $\mathbf{r}_B$ though. Nevertheless, the following straight line:
\begin{equation}\label{eq:lambda}
\lambda \hat{\mathbf{n}}+\mathbf{r}_B
\end{equation}
will pass through $\mathbf{r}_B$ when $\lambda=0$ and, being parallel to the plane's normal, it will pierce through the plane spanning the smallest distance among the point and the plane. Let's parameterize that special value for $\lambda$ as $\lambda_0$. Then, Eqn.~\eqref{eq:plane} turns more specific (because the intersection of a plane and a straight line is a point), and it helps us determine $\lambda_0$ explicitly. Indeed, substituting for $\mathbf{r}$ the equation of the straight line, Eqn.~\eqref{eq:lambda} for the value of $\lambda=\lambda_0$ for which the line pierces the plane, we get:
\begin{equation}
\left[ \lambda_0 \hat{\mathbf{n}}+\mathbf{r}_B -(\mathbf{r}_{A_1}\cdot \hat{\mathbf{n}}) \hat{\mathbf{n}} \right]\cdot \hat{\mathbf{n}}=0,
\end{equation}
which can be simplified into
\begin{equation}\label{eq:lambda0}
\lambda_0=(\mathbf{r}_{A_1}-\mathbf{r}_{B})\cdot \hat{\mathbf{n}}.
\end{equation}

This way, plugging the value of $\lambda_0$ found in Eqn.~\eqref{eq:lambda0} into Eqn.~\ref{eq:lambda}, the closest point to $\mathbf{r}_B$ within the plane (which we will call $\mathbf{r}_0$) ends up being given by:
\begin{equation}
\mathbf{r}_0=\left[ (\mathbf{r}_{A_1}-\mathbf{r}_{B})\cdot\hat{\mathbf{n}}\right]\hat{\mathbf{n}}+\mathbf{r}_B,
\end{equation}
so the shortest vector between point $\mathbf{r}_{B}$ and $\mathbf{r}_0$ ({\em i.e.}, $\mathbf{r}_{B}-\mathbf{r}_0$) ends up being
\begin{equation}
\mathbf{r}_{B}-\mathbf{r}_0=\left[(\mathbf{r}_{B}-\mathbf{r}_{A_1}) \cdot\hat{\mathbf{n}}\right]\hat{\mathbf{n}}.
\end{equation}

$\Delta_z$ is the projection of $\mathbf{r}_{B}-\mathbf{r}_0$ into the local normal $\hat{\mathbf{n}}$:
\begin{equation}\label{eq:updown}
\Delta_z=(\mathbf{r}_{B}-\mathbf{r}_0)\cdot\hat{\mathbf{n}}= (\mathbf{r}_{B}-\mathbf{r}_{A_1})\cdot\hat{\mathbf{n}},
\end{equation}
where $\hat{\mathbf{n}}\cdot \hat{\mathbf{n}}=1$ was employed. Eqn.~\eqref{eq:updown} formalizes the concept of ``up'' or ``down'' for every $B$ atom within the supercell, at any temperature, and at any time, as long as the supercell does not turn amorphous.

\section{Band broadening arising from deformation potential}\label{sec:defpot}

\begin{table}
\caption{Average and standard deviation of deformation potential, $E_s$, for low-buckled silicene, germanene, and stanene.}
\label{table:Es}
\begin{tabular}{c c c c}
\hline
\hline
Material & Temperature (K) & $ \bar{E}_s$ (eV) &  $\sigma_{E_s}$ (eV)\\
\hline
\hline
silicene  & 200 & 0.0001 & 0.0009 \\
silicene  & 550 & 0.0002 & 0.0026 \\
silicene  & 600 & 0.0000 & 0.0060 \\
silicene  & 650 & 0.0001 & 0.0052 \\
\hline
germanene & 100 & 0.0000 & 0.0012\\
germanene & 500 & 0.0000 & 0.0022\\
germanene & 600 & 0.0000 & 0.0025\\
germanene & 650 & 0.0000 & 0.0020\\
\hline
stanene   & 100 & 0.0000 & 0.0010\\
stanene   & 300 & 0.0000 & 0.0013\\
\hline
\hline
\end{tabular}
\end{table}

We follow the process introduced in Ref.~\cite{puddles2} to estimate band broadening due to finite-temperature atomistic motion. Here, we use one hundred snapshots, equally separated, and create a primitive unit cell using the average lattice constant and buckling height per snapshot. The distance among three neighboring atoms on each of these average primitive unit cells is called $d$.

We also calculate the $\pi-$electron hopping integral $t$ at zero temperature, by fitting the linear dispersion of these materials away from the gap (they all obey a Dirac Equation for fermions with mass). In that linear region, the electron's velocity (known as Fermi velocity $v_F$) is given by $v_F=\frac{\Delta E}{\hbar\Delta k}$, and the hopping integral $t$ is
\begin{equation}
t=\frac{2}{\sqrt{3}a_{0,lb}}\frac{\Delta E}{\Delta k},
\end{equation}
with $a_{0,lb}$ the zero-temperature lattice constant from Table \ref{table:lb_and_planar}. Furthermore, $d_0=\sqrt{\left(\frac{a_{0,lb}}{\sqrt{3}}\right)^2+\Delta_{z,lb}^2}$.

With this procedure, we found $t = 1.000$ eV for low-buckled silicene, t=0.954 eV for low-buckled germanene, and t=0.744 eV for low-buckled stanene, and the deformation potential $E_s$ is estimated as a linear deviation related to the average change in interatomic distances on the unit cell:
\begin{equation}\label{eq:Es}
E_s=-t\frac{d-d_{0}}{d_{0}},
\end{equation}
which is adapted from Ref.~\cite{puddles2}. A reduction of the size of the unit cell makes $d<0$ and increases the electron cloud density, which results on electron repulsion and a local upshift of the chemical potential $E_s$ [Eqn.~\eqref{eq:Es}], for an electron-deficient region. Alternately, an increase of the size of the unit cell makes $d>0$ and decreases the local electron density, which results on a downshift of $E_s$, and an electron-rich region. See Refs.~\cite{puddles1,puddles2,puddles3} for more details.

On a real material, there are electron-rich and electron-rich regions [see, e.g., Refs.~\cite{puddles0,puddles1,puddles2,puddles3,puddles4}], which make the local conduction and valence band fluctuate with respect to the midgap. Assuming ergodic behavior, we calculate the electronic structure on a given {\em average primitive unit cell}, and extract the band lifetime and the effects of electron-phonon coupling at finite temperature as listed in Table \ref{table:Es}.


%

\end{document}